\documentclass[
journal=largetwo,
manuscript=article-type,
year=2020,
volume=37,
]{cup-journal}

\usepackage{amssymb, amsmath}
\usepackage[nopatch]{microtype}
\usepackage{booktabs}
\usepackage{multirow}
\usepackage{hyperref}

\makeatletter
\newcommand*{\citelinktext}[2]{%
	\hyper@@link[cite]{}{cite.0@#1}{#2}}
\makeatother

\title{Less Wound and More Asymmetric: JWST Confirms the Evolution of Spiral Structure in Galaxies at $z \lesssim 3$}

\author{Ilia V. Chugunov}
\affiliation{Pulkovo Astronomical Observatory, Russian Academy of Sciences, St. Petersburg 196140, Russia}
\email[I. V. Chugunov]{chugunov21@list.ru}

\author{Alexander A. Marchuk}
\affiliation{Pulkovo Astronomical Observatory, Russian Academy of Sciences, St. Petersburg 196140, Russia}
\alsoaffiliation{St. Petersburg State University, Universitetskii pr. 28, St. Petersburg 198504, Russia}

\author{Aleksandr V. Mosenkov}
\affiliation{Department of Physics and Astronomy, N283 ESC, Brigham Young University, Provo, UT 84602, USA}

 %% First letter not capped

\begin{document}
	
	\begin{abstract}
		Spiral galaxies are ubiquitous in the local Universe. However the properties of spiral arms in them are still not well studied, and there is even less information concerning spiral structure in distant galaxies. We aim to measure the most general parameters of spiral arms in remote galaxies and trace their changes with redshift. We perform photometric decomposition, including spiral arms, for 159 galaxies from the HST COSMOS and JWST CEERS and JADES surveys, which are imaged in optical and near-infrared rest-frame wavelengths. We confirm that, in our representative sample of spiral galaxies, the pitch angles increase, and the azimuthal lengths decrease with increasing redshift, implying that the spiral structure becomes more tightly wound over time. For the spiral-to-total luminosity ratio and the spiral width-to-disc scale length ratio, we find that band-shifting effects can be as significant as, or even stronger than, evolutionary effects. Additionally, we find that spiral structure becomes more asymmetric at higher redshifts.
	\end{abstract} 
	
	\section{Introduction}
	Galaxies evolve over cosmic time~\citep{Conselice2014}, and with the aid of the deepest available surveys, some aspects of galaxy evolution have been explored at redshifts as early as $z = 8$ (see, e.g.~\citealp{Conselice2016,Ormerod2024}). However, in the high-$z$ domain, only the most general characteristics of galaxies, such as size, shape, or number density, can be studied. The evolution of galaxies in terms of more specific properties has been investigated in a narrower range of redshifts. For example, galaxies generally tend to be clumpier and less symmetric at higher redshifts, at least up to $z \approx 3$~\citep{Margalef-Bentabol2022}. The evolution of the bar fraction has also been studied for $z \lesssim 3$~\citep{Guo2023,LeConte2023}, showing a rapid decrease with increasing $z$, albeit the exact form of the trend depends on galaxy mass. Bulge mass also decreases with increasing $z$, as seen in both observations (e.g.~\citealp{Sachdeva2017}) and numerical simulations (e.g.~\citealp{Brooks2016}). \citet{Bruce2014} also demonstrated that the fraction of bulge-dominated galaxies increases as $z$ decreases. Indeed, modern cosmology predicts bulge growth over time due to mergers and dynamical evolution~\citep{Hopkins2010}. Despite this, our current understanding of the evolution of spiral structure remains rather limited.
	
	The majority of luminous galaxies (75\% for $M(B)<-20$~mag) in the local Universe exhibit a spiral pattern~\citep{Conselice2006}. Despite a long observational history (dating back to~\citealp{Rosse1850}), the questions of the origin and evolution of spiral arms in galaxies still remain unsettled. Several theories aim to explain the nature of spiral arms, none of which have been definitively ruled out by observations~\citep{Dobbs2014}. Among the most notable theories, the first is the density wave theory proposed by~\citealp{Lin1964} (for a modern review, see~\citealp{Shu2016}). In this framework, the spiral structure is a feature supported by a long-lived, quasi-stationary density wave in the galactic disc. The second is the dynamic spirals theory, which suggests that spiral structures consist of small elements that constantly form from instabilities and disappear after a short time. Observational tests can, in principle, be interpreted as evidence supporting one or another proposed theory. For instance, the Pringle–Dobbs test~\citep{Pringle2019} links the distribution of the cotangent of a pitch angle to the main mechanism of spiral arm formation.
	
	There are numerous studies devoted to the spiral structure of nearby galaxies (see discussion and references in~\citealp{Sellwood2022a}). For distant galaxies, systematic works concerning specifically the spiral structure are scarce and focus mostly on its pitch angle. In particular, \citet{Davis2012} found that pitch angle does not change with time, but~\citet{Reshetnikov2022, Reshetnikov2023} report that pitch angle increases with redshift up to $z \approx 1$. Other studies on this subject include \citet{Savchenko2011} and \citet{Martinez-Garcia2023}. At the same time, it is known that spiral galaxies are abundant at high redshifts. Spiral galaxy formation is believed to occur mostly at $1.5 < z < 3$, because the fraction of spiral systems drops with increasing $z$ over this range~\citep{Margalef-Bentabol2022}, but even more distant examples are known, up to $z = 4.41$~\citep{Tsukui2021}, and the fraction of spiral galaxies is thought to remain significant at least up to $z \approx 4$~\citep{Kuhn2023}.
	
	The most widely used method for measuring the pitch angle of the spiral arms is 2D Fourier analysis and its variations (see e.g.~\citealp{Davis2012, Yu2018a, Diaz-Garcia2019}). Fourier-based methods are popular for various reasons. In particular, they are relatively easy to implement, they do not rely on assumptions concerning the parameters of the spiral arms, and require minimal human control. However, these methods by themselves are hardly capable of measuring parameters other than pitch angle and are unable to determine them for an individual spiral arm.
	
	Examining the light distribution along multiple slices of a spiral arm is a way to retrieve various properties of spiral arms, not only pitch angles. This method is non-parametric, and~\citet{Savchenko2020} have shown that pitch angles, arm widths, and the contribution of the spiral arms to the overall luminosity of a galaxy can be obtained with it in practice. Moreover, even corotation radii can be determined using this method~\citep{Marchuk2024a}. \citet{Mosenkov2023, Mosenkov2024} have also implemented this method to trace the spiral structure up to the outermost parts of galaxies using deep photometry. Potentially, photometric decomposition is an even more powerful tool for measuring properties of spiral arms, because it uses the overall light distribution in a galaxy. Nowadays, however, it is predominantly used while ignoring the spiral arms, mostly because of their diverse appearance and complex structure, and the lack of established photometric models for spiral arms. Nevertheless, some advances in this area have been made. For example, the \verb|GALFIT| package for photometric decomposition~\citep{Peng2010} allows one to model various non-axisymmetric features, including spirals, with Fourier and bending modes. Some studies have followed this approach to model spiral arms, e.g.~\citet{Lasker2014} and \citet{Gao2017}. A simple model designed specifically to reproduce spiral arms was used in the decomposition by~\citet{Lingard2020}. Recently, we developed and applied a more versatile model of spiral arms, capable of capturing a wider range of features, presented in~\citet{Chugunov2024} and \citet{Marchuk2024b}, hereafter \citelinktext{Chugunov2024}{Paper I} and \citelinktext{Marchuk2024b}{Paper II}, respectively. Studies that measure the properties of spiral arms with the aid of photometric decomposition are even less common. \citet{Davis2012} has measured pitch angles of spiral structure in galaxies using Fourier transform decomposition. In our \citelinktext{Chugunov2024}{Paper I}, we examined various properties of the spiral arms for 29 local galaxies in the 3.6~$\mu$m band and explored different relationships between the parameters of the spiral structure and other galaxy characteristics.
	
	The primary goal of this paper is to explore the various properties of spiral arms in {\it remote} galaxies in order to determine how they evolve with cosmic time, using the most advanced observations from HST and JWST. Specifically, we aim to measure parameters such as pitch angle, arm width, arm length, and the contribution of spiral arms to the overall luminosity of galaxies at different redshifts. By analysing these properties, we seek to understand how the spiral structure changes from the early Universe to the present day, offering insights into the processes driving the formation and evolution of spiral structure in galaxies.
	
	Our paper is organised as follows. In Section~\ref{sec:data}, we describe the sample of galaxies and images used. In Section~\ref{sec:model}, we present the spiral arm model and provide some details of our decomposition. We describe the results and test their validity in Sections~\ref{sec:results} and~\ref{sec:validation}, respectively. Our results are discussed in Section~\ref{sec:discussion}. We summarise our main findings and draw final conclusions in Section~\ref{sec:conclusions}.
	
	\label{sec:cosmology}
	In this work, we adopt a standard flat $\Lambda$CDM cosmology with $\Omega_m = 0.3$, $\Omega_\Lambda = 0.7$, and $H_0$~=~70~km~s$^{-1}$~Mpc$^{-1}$. We use Cosmology Calculator by~\citet{Wright2006} to perform all relevant calculations. We express all magnitudes in the AB-system~\citep{Oke1983}.
	
	\section{Data}
	\label{sec:data}
	To investigate spiral structure in remote galaxies, we use data from the COSMOS survey by HST~\citep{Koekemoer2007} and from two JWST surveys, namely CEERS~\citep{Yang2021, Bagley2023} and JADES~\citep{Eisenstein2023, Rieke2023}. We use images of galaxies with a recognisable spiral structure, which naturally requires selecting galaxies that are both bright and large enough for clear observation.
	
	In short, our final sample consists of 126 galaxies from the COSMOS survey at $0.10 \leq z \leq 1.02$, imaged using a single F814W filter. These galaxies were visually selected based on the catalogue by~\citet{Reshetnikov2023}. Additionally, 22 galaxies were taken from the CEERS survey and 11 from JADES, covering a combined redshift range of $1.05 \leq z \leq 3.30$. These galaxies were observed in multiple NIRCam bands, and their selection was primarily based on visual inspection of the images.
	
	For the COSMOS survey, deep images in the F814W filter are available for an area of almost 2 deg$^2$, with a pixel size of 0.03 arcseconds. Both JWST surveys offer even deeper images in multiple infrared NIRCam filters with the same pixel size. These two surveys have slightly different sets of available filters, and in total, they cover the 0.9--4.4 $\mu$m range. The fields captured within CEERS and JADES have areas of nearly 100 arcmin$^2$ and 175 arcmin$^2$, respectively. In Table~\ref{tab:bands}, we provide the list of bands used in this paper, along with the characteristics of the images: pivot wavelength $\lambda_p$, full-width at half-maximum (FWHM) of their point spread function (PSF), and the photometric depth of the corresponding surveys.
	
	\begin{table}[hbt!]
		\begin{threeparttable}
			\caption[]{Photometric bands used and their parameters}
			\label{tab:bands}
			\begin{tabular}{ccccc}
				\toprule
				\headrow Band & $\lambda_p$ & PSF FWHM & \multicolumn{2}{c}{Photometric depth}\\
				\headrow & $\mu$m & arcsec & \multicolumn{2}{c}{mag/arcsec$^{-2}$}\\
				\midrule
				\headrow \multicolumn{3}{c}{HST filters} & \multicolumn{2}{c}{COSMOS}\\
				\bottomrule
				F814W & 0.806 & 0.095 &  \multicolumn{2}{c}{29.8}\\
				\toprule
				\headrow \multicolumn{3}{c}{JWST filters} & CEERS & JADES \\
				\bottomrule
				F090W & 0.901 & 0.033 &  & 31.5 \\
				F115W & 1.154 & 0.040 & 31.1 & 31.6 \\
				F150W & 1.501 & 0.050 & 30.9 & 31.8 \\
				F182M & 1.845 & 0.062 &  & 31.0 \\
				F200W & 1.990 & 0.066 & 31.2 & 31.9 \\
				F210M & 2.093 & 0.071 &  & 30.6 \\
				F277W & 2.786 & 0.092 & 32.0 & 32.6 \\
				F335M & 3.365 & 0.111 &  & 32.5 \\
				F356W & 3.563 & 0.116 & 32.0 & 32.6 \\
				F410M & 4.092 & 0.137 & 31.2 & 31.8 \\
				F444W & 4.421 & 0.145 & 31.6 & 31.8 \\
				\bottomrule
			\end{tabular}
			\begin{tablenotes}[hang]
				\item[]Data from~\citet{Koekemoer2007, Bagley2023, Rieke2023}. Photometric depth was estimated directly from images and calculated as 3$\sigma$ fluctuation within a set of 1000 randomly oriented $10\times10$ arcsec$^2$ boxes.
			\end{tablenotes}
		\end{threeparttable}
	\end{table}
	
	Part of our sample taken from the COSMOS survey is based on that presented in~\citet{Reshetnikov2023}. The objects were selected from the sample of 26113 bright (apparent magnitude $m_\text{F814W} < 22.5$ mag) galaxies in the COSMOS field from~\citet{Mandelbaum2012}. Firstly, the apparent axis ratio ($b/a$) was determined for each galaxy using the SExtractor package~\citep{Bertin1996}, and galaxies with $b/a > 0.7$ were selected. Based on careful visual inspection, a sample was then selected consisting of 171 galaxies with a pronounced spiral structure, which constitute the final sample of~\citet{Reshetnikov2023}. In this work, we managed to obtain decomposition results (see Section~\ref{sec:decomposition})  for only 126 galaxies from this sample. For others, we were unable to produce a plausible fit, often due to structural irregularities, and these galaxies will not be considered further. The sample objects were identified with the COSMOS2020 catalogue~\citep{Weaver2022}. For each galaxy, we adopted the photometric redshift from COSMOS2020 or, where it was unavailable, from~\citet{Ilbert2009}. If brightest galaxies ($m_\text{F814W} < 22.5$ mag, similar to our criterion for galaxies selection) are considered, then the fraction of cases where photometric redshift $z_\text{phot}$ differs significantly from spectroscopic $z_\text{spec}$ (by more than $0.15 \times (1 + z_\text{spec})$, which is a standard criterion), is less than 1\% in both~\citet{Weaver2022} and~\citet{Ilbert2009}.
	
	Both remaining CEERS and JADES subsamples are new, consisting of 22 and 11 galaxies, respectively. Because the subsample from COSMOS already covers redshifts $z \lesssim 1$ sufficiently and the JWST data is deeper, only galaxies with $z > 1$ were selected from the CEERS and JADES data. These subsamples contain some spiral galaxies from~\citet{Costantin2023, Guo2023, LeConte2023, Kalita2023}, but most galaxies were added based on visual inspection of the CEERS and JADES fields and selection of galaxies with prominent spiral structure. Similarly to the COSMOS subsample, we initially selected more galaxies than those included in the final sample (29 from CEERS and 15 from JADES), but we accepted decomposition results for only 33 objects combined. For CEERS, we identified objects based on their coordinates from the ~\citet{Stefanon2017} catalogue and used the photometric redshifts provided in it. For JADES, object identification and photometric redshift measurements were conducted as a part of the survey~\citep{Rieke2023}. For both sources, nearly 5\% of the galaxies have a large discrepancy, between the photometric and spectroscopic redshifts (defined as more than $0.15 \times (1 + z_\text{spec})$ in ~\citealp{Rieke2023} and $5\sigma = 0.10 \times (1 + z_\text{spec})$ difference in ~\citealp{Stefanon2017}). Note that, similar to the COSMOS sample, the photometric redshift accuracy for galaxies in our study is probably higher than the average, because they are brighter than average galaxies from the mentioned works.
	
\begin{figure}[!ht]
		\centering
		\includegraphics[width=0.99\linewidth]{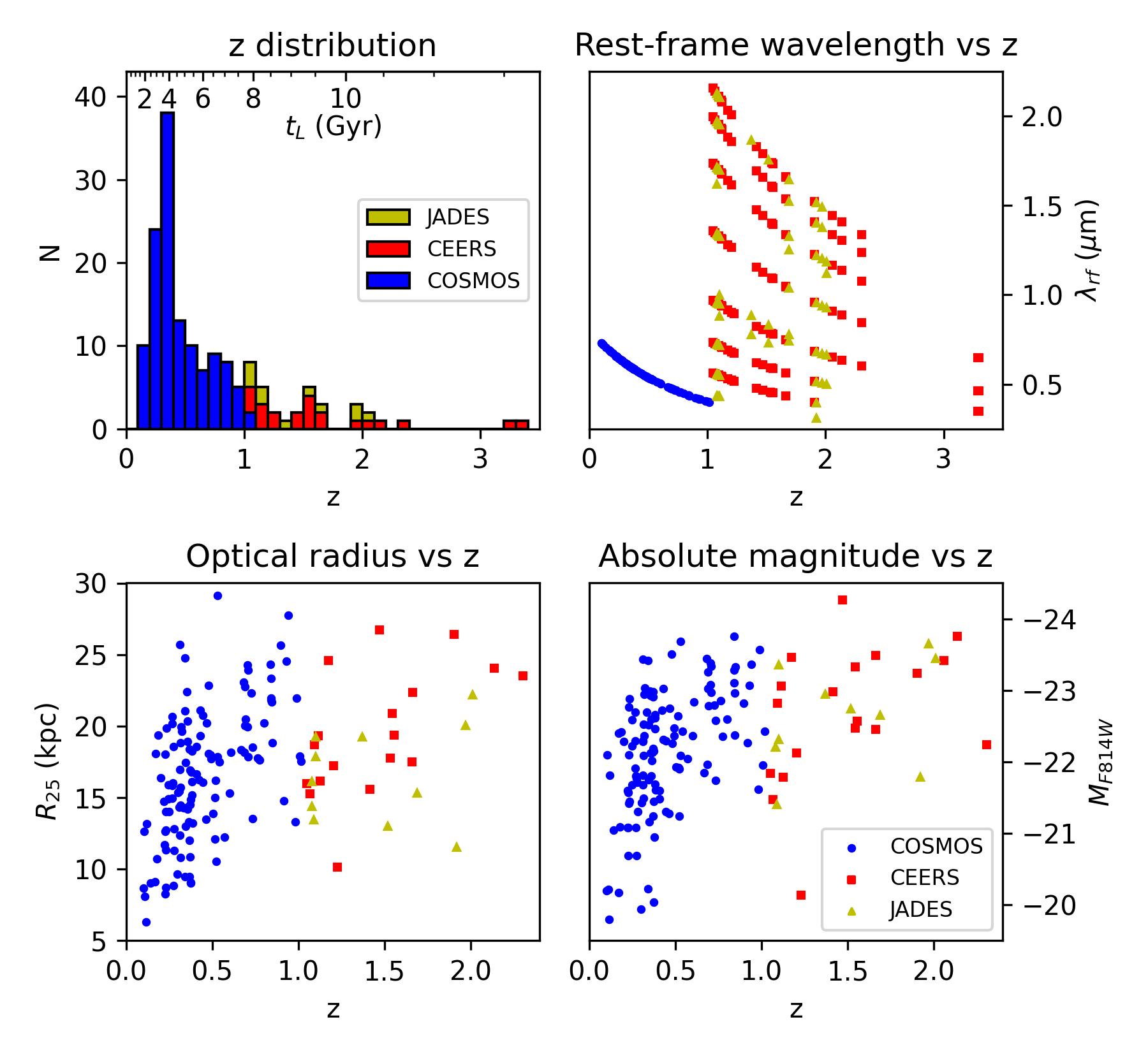}
		\caption{General parameters of galaxies from our sample. Top left: the distribution of sample galaxies by redshift $z$. The top axis shows the lookback time $t_L$ corresponding to $z$. Top right: each point represents a single image, with the rest-frame wavelength $\lambda_\text{rf}$ displayed versus $z$. Bottom left: disc optical radius in the rest-frame is charted versus $z$ (for CEERS and JADES, it is measured in the F200W or F210M band, depending on availability). Bottom right: absolute magnitude corresponding to the rest-frame F814W band versus $z$.}
		\label{fig:basic_data}
	\end{figure}

	Overall, our sample contains 159 galaxies with a prominent spiral structure at redshifts $0.1 \leq z \leq 3.3$, with 329 individual images from the HST and JWST. Part of the sample from COSMOS contains 126 galaxies with lower $z$ (the lowest is $z = 0.10$, the greatest is $z = 1.02$). Of the remaining 33 galaxies from JWST, the lowest redshift is $z = 1.05$ and the highest is $z = 3.30$, although almost all galaxies except the two most distant ones have $z < 2.31$. Thanks to the multi-band data available, we have analysed 203 individual images from JWST. Note that the set of available NIRCam filters is not the same for all galaxies; for each galaxy, at least 3 filters were used. Therefore, the JWST subsample allows us to significantly extend our $z$ coverage and also analyse how the properties of spiral arms vary with wavelength. The general properties of galaxies in our sample are shown in Figure~\ref{fig:basic_data}.
	
	The rest-frame wavelength $\lambda_\text{rf}$ is calculated as $\lambda_\text{rf} = \lambda_p / (1 + z)$, where $\lambda_p$ is the filter pivot wavelength (see Table~\ref{tab:bands}). We derived the rest-frame absolute magnitudes from the total galaxy fluxes extracted through our decomposition models, incorporating galaxy redshift to account for distance and cosmological dimming. We also applied a correction for galactic extinction in the corresponding fields using the~\citet{Schlafly2011} extinction map. However, since we use filters with infrared $\lambda_p$, the extinction does not exceed 0.03 for any image. Subsequently, the magnitudes were recalculated from the rest-frame to the wavelength of the F814W filter $\lambda_\text{F814W}$. For galaxies from the COSMOS survey, we used the multiwavelength photometry provided by~\citet{Weaver2022}, which includes a large number of filters from UV to mid-IR and, therefore, allows the calculation of K-corrections for each object. To calculate K-corrections for CEERS and JADES galaxies, we use the distribution of absolute magnitudes in different rest-frames which were derived from total fluxes of the best-fitting decomposition model. In both cases, we interpolate between magnitudes $m$ as a function of $\lambda_p$ and find the interpolated magnitude at $\lambda_\text{F814W} (1 + z)$ which translates into $\lambda_\text{F814W}$ in the rest-frame of a galaxy. Therefore, the difference $m(\lambda_\text{F814W} (1 + z)) - m(\lambda_p)$ is effectively the K-correction needed to translate magnitude in filter with $\lambda_p$ pivot wavelength to F814W wavelength.
	
	We also define subsamples of galaxies based on their parameters, which we will use later when examining various relationships. First, we consider a sample of bright galaxies, down to some magnitude to exclude faintest galaxies that are observed only at low redshifts. Specifically, we take $M_\text{F814W} = -22.5$ as the limiting absolute magnitude because galaxies with such magnitude are represented at all studied redshifts (see Figure~\ref{fig:basic_data}), and there are 63 galaxies brighter than this limit. Additionally, we distinguish a sample of 93 two-armed spiral galaxies and a sample of 52 barred galaxies, both determined visually. The reasons for considering these subsamples are 1) that grand-design spirals (which are by definition a distinct type of spiral structure) are exclusively two-armed~\citep{Elmegreen1987}, and 2) that there is a known connection between bars and spirals~\citep{Athanassoula2010}. 
	
	Overall, the examined sample is among the biggest compared to previous works on decomposition with the spiral arms (e.g., \citealp{Lingard2020}), and contains a significant fraction of observed spiral galaxies at large $z$ at all (as estimated in \citealp{Kuhn2023} using JWST data).
	
	\section{Spiral arm model}
	\label{sec:model}
	To model the 2D light distribution in spiral arms, we use a function with 21 parameters for each individual arm. This model is adapted from \citelinktext{Chugunov2024}{Paper I}, with minor, mostly technical changes. The core principles and motivations remain the same, and we refer the reader to Section 3 in \citelinktext{Chugunov2024}{Paper I}. The most comprehensive description of the model and the changes compared to the version used in \citelinktext{Chugunov2024}{Paper I} can be found at~\url{https://github.com/IVChugunov/IMFIT_spirals}. This repository also contains the modified \verb|IMFIT| package~\citep{Erwin2015} for photometric decomposition with this function implemented (see Section~\ref{sec:decomposition}), and it is available for anyone to use.
	
	Formally, in our model, the surface brightness $I$ is a function of polar coordinates $(r, \varphi)$. We consider the origin to be at the centre of the galaxy under consideration, and the coordinate plane matches the galactic plane. Four parameters are used to define the galactic galaxy plane: the coordinates of the centre (X$_0$, Y$_0$) and the plane orientation parameters --- galaxy inclination $i$ and position angle (PA).
	
	Let the beginning of a spiral arm be located at the azimuthal angle $\varphi_0$. For convenience, we will use the coordinate $\psi$ (winding angle), which is measured from the beginning of a spiral arm in the direction of its winding. Thus, $\psi = \varphi - \varphi_0$ if the spiral arm winds counterclockwise, and $\psi = \varphi_0 - \varphi$ if it winds clockwise. The 2D distribution is then presented in the following generalised way:
	
	\begin{equation}
		I(r, \psi) = I_\parallel(r(\psi), \psi) \times I_\bot(r - r(\psi), \psi)\,.
	\end{equation}
	
	Here, $I_\parallel$ is the surface brightness distribution along the ridge line of the spiral, which, in turn, is described by the shape function of the arm $r(\psi)$. $I_\bot$ is the surface brightness distribution across the arm. We will now consider each of these functions separately.
	
	\subsection{Arm shape}
	The shape function of an arm, $r(\psi)$, describes the overall shape of the spiral arm and, more precisely, defines its ridge line. In the case of a logarithmic spiral, $\log r$ is a linear function of $\psi$, meaning the spiral arm has a constant pitch angle $\mu$. However, it is known that real galaxies often exhibit pitch angles that vary with radius~\citep{Savchenko2013}. Therefore, in this model, $\log r$ is represented as a 4th-degree polynomial of $\psi$, allowing the pitch angle to be variable as a 3rd-degree polynomial of $\psi$.
	
	\begin{equation}
		\label{eq:r_psi}
		r(\psi) = r_0 \times \exp \left(\sum_{n=1}^4 k_n (\psi / \psi_\text{end})^n\right)\,.
	\end{equation}
	
	Coefficients $k_n$ are convenient for presenting Equation~\ref{eq:r_psi}, but using them directly as function parameters makes it more difficult to control the model, and it also increases the likelihood of parameter degeneracy. Instead, we employ more meaningful input parameters that, in turn, define the coefficients $k_n$. Among these input parameters are the radii marking the beginning and end of the arm ($r_0$ and $r_\text{end}$, respectively). The input parameters also include the coefficients $m_{2 \ldots 4}$, which have a geometric interpretation as deviations from the logarithmic spiral shape. These coefficients correspond to specific linear combinations of $k_{1 \ldots 4}$ and represent the Legendre polynomial expansion coefficients for the pitch angle.
	
	Strictly speaking, the $k_n$ coefficients are derived from the input parameters in the following way:
	
	\begin{equation}
		\label{eq:k_m}
		\left\{\begin{array}{l}
			m_1 = \ln{(r_\text{end} / r_0)}\\
			k_1 = m_1 - m_2 + m_3 - m_4 \\
			k_2 = m_2 - 3 m_3 + 6 m_4 \\
			k_3 = 2 m_3 - 10 m_4 \\
			k_4 = 5 m_4
		\end{array}\right.\
	\end{equation}
	
	\subsection{Distribution of the light along the arm}
	The function $I_\parallel$ describes the light distribution along the ridge line of the spiral arm, defined by $r(\psi)$. Because the spiral arm is part of the disc, it is natural to assume that the light distribution is exponential with radius along the spiral arm. To ensure the beginning and the end of the arm are smooth, this exponential decrease is multiplied by a truncation function of $\psi$. The main part describing the exponential decline, $I_{r \parallel}$, and the modification function, $I_{\psi \parallel}$, can be separated and considered independently:
	
	\begin{equation}
		I_\parallel(r, \psi) = I^\text{sp}_0 \times I_{r \parallel}(r) \times I_{\psi \parallel}(\psi)\,,
	\end{equation}
	
	\begin{equation}
		I_{r \parallel}(r) = e^{-r/h_s}\,,
	\end{equation}
	
	\begin{equation}
		I_{\psi \parallel}(\psi) =
		\left\{\begin{array}{ll}
			3 \left(\frac{\psi}{\psi_\text{gr}}\right)^2 - 2 \left(\frac{\psi}{\psi_\text{gr}}\right)^3, & 0 \leq \psi < \psi_\text{gr}\\
			1, & \psi_\text{gr} \leq \psi < \psi_\text{cut}\\
			\frac{\psi_\text{end} - \psi}{\psi_\text{end} - \psi_\text{cut}}, & \psi_\text{cut} \leq \psi \leq \psi_\text{end}\,.
		\end{array}\right.\
	\end{equation}
	
	Above, $I^\text{sp}_0$ is the projected surface brightness from the exponential part at the galaxy centre, and $h_s$ is the exponential scale of the spiral arm; thus, $I_{r \parallel}$ is similar in form to the exponential disc function. The modification function is designed to keep the overall function exponential for most values of $\psi$ (hence, part of $I_{\psi \parallel}$ is constantly 1). The parameters $\psi_\text{gr}$ and $\psi_\text{cut}$ define the beginning and the end of the purely exponential part, while outside this range, there is a smooth transition to zero brightness at $\psi = 0$ and $\psi = \psi_\text{end}$, provided by $I_{\psi \parallel}(0) = I_{\psi \parallel}(\psi_\text{end}) = 0$.
	
	\subsection{Distribution of the light across the arm}
	
	The distribution of the light across the arm is defined by the function $I_\bot$ which in its form is very similar to the asymmetric S\'ersic function. As spiral arm width can vary, $I_\bot$ also has to depend on the position along the arm. Therefore, we express it as a function of two parameters: $\rho = r - r(\psi)$ and $\psi$. Effectively, $\rho$ is the radial distance from the point to the ridge line of the spiral arm.
	
	\begin{equation}
		\label{eq:I_bot}
		I_\bot^\text{in/out}(\rho, \psi) = \exp \left(-\ln(2) \times \left( \frac{|\rho|}{w_\text{loc}^\text{in/out}(\psi)} \right)^{\frac{1}{n^\text{in/out}}} \right)\,,
	\end{equation}
	
	where $w_\text{loc}^\text{in/out}(\psi)$ is the half-width at half-maximum (HWHM) of the arm in the radial direction, the dependence of which on $\psi$ is described below (Eq.~\ref{eq:w_loc}), $n^\text{in/out}$ is the S\'ersic index that defines the light distribution within the arm.
	
	The indices $^\text{in}$ and $^\text{out}$ denote the inner part of the arm relative to the ridge-line of the arm ($\rho < 0$), and the outer part ($\rho > 0$), respectively. The inner and outer S\'ersic indices, $n^\text{in}$ and $n^\text{out}$, are independent of each other, as are the widths $w_\text{loc}^\text{in}$ and $w_\text{loc}^\text{out}$.
	
	\begin{equation}
		\label{eq:w_loc}
		w_\text{loc}^\text{in/out}(\psi) = w_r \frac{1 \mp S}{2} \times \exp\left(\gamma_\text{in/out} \left(\frac{\psi}{\psi_\text{end}} - 0.5\right)\right)\,.
	\end{equation}
	
	The average radial width $w_r$ is the local width at the midpoint of the arm in terms of $\psi$, formally defined as $w_r = w_\text{loc}(\psi_\text{end} / 2)$. The ``true'' perpendicular width is slightly smaller than the radial one, and to maintain strict notation, we define it as $w = w_r \times \cos \mu$, where $\mu$ is a pitch angle of the arm calculated from its shape. Skewness $S$ measures how the outer part is wider than the inner part. Finally, $\gamma^\text{in/out}$ describes the rate of increase or decrease in the spiral arm width.
	
	\subsection{Photometric decomposition}
	\label{sec:decomposition}
	To perform photometric decomposition of galaxy images, we use the \verb|IMFIT| package~\citep{Erwin2015} which we have modified to implement our spiral function (available at~\url{https://github.com/IVChugunov/IMFIT_spirals}).
	
	For each galaxy, we employ a model with several appropriate components, which may include a disc (always present), a bulge (present in most cases, except for 6 objects), a bar (adopted for 52 galaxies), and, in individual cases, a ring (included only for 5 galaxies). A suitable number of spiral arms is also added to the model; the necessity of each component and the number of spiral arms are determined visually. In particular, the inclusion of a galaxy in the subsamples of barred and/or two-armed galaxies (described in Section~\ref{sec:data}) corresponds to the presence of a bar and the number of spiral arms in the model as determined at this stage.
	
	If present, each component is modelled with a function described in the main \verb|IMFIT| paper~\citep{Erwin2015}. For discs, we employ an exponential function of radius~\citep{Freeman1970}, while bulges and bars are described with a S\'ersic function~\citep{Sersic1968}. However, the isophotes of bars have a generalised ellipse shape instead of a pure ellipse (\citealp{Peng2002}; see also the function \verb|Sersic_GenEllipse| in \verb|IMFIT|). Finally, rings are modelled with a Gaussian radial profile centred at a certain radius (\verb|GaussianRing| in \verb|IMFIT|).
	
	Needless to say, a proper photometric decomposition must account for the PSF and pixel weights, and, if necessary, some parts of the image must be masked (e.g., other galaxies), which can be readily implemented in \verb|IMFIT|. The PSF for the HST F814W filter was obtained using the \verb|TinyTim| code~\citep{Krist2011}, whereas for all JWST filters, the \verb|WebbPSF| code was used, as described in~\citet{Perrin2014}. Error maps, which determine the pixel weights, are provided along with images for the CEERS and JADES surveys. For COSMOS, instrumental maps are unavailable, so we constructed error maps using only galaxy images, using the method described in \citelinktext{Marchuk2024b}{Paper II}. For all galaxies, the masks were prepared manually based on visual analysis of the images, and we usually masked only external objects relative to the galaxy. In some cases, when the initial decomposition was completed, it turned out that some areas of the galaxy, like bright clumps, significantly affected the results (most often causing the best-fitting spiral arm to change its shape to encompass the bright area), and such areas were masked out for the final decomposition. If images in different filters are present for a single galaxy, in most cases, the mask was the same between them.
	
	The main difficulty in photometric decomposition with such a complex model is finding a proper initial guess for the fit parameters. To address this, we created a Python script that computes an initial guess from manually marked spiral arms (similar to the marking method used in~\citealp{Reshetnikov2022} for the COSMOS subsample) and combines it with the results of the decomposition using only the ``classical'' components. Inclinations of galaxies were also determined from this ``classical'' decomposition (most of the galaxies are close to face-on orientation; exact values can be found in online material, see Section~\ref{sec:results}). This approach allowed us to make the decomposition process as automatic as possible and obtain results for a much larger number of galaxies than in \citelinktext{Chugunov2024}{Paper I}.
	
\begin{figure}[!ht]
		\centering
		\includegraphics[width=0.99\linewidth]{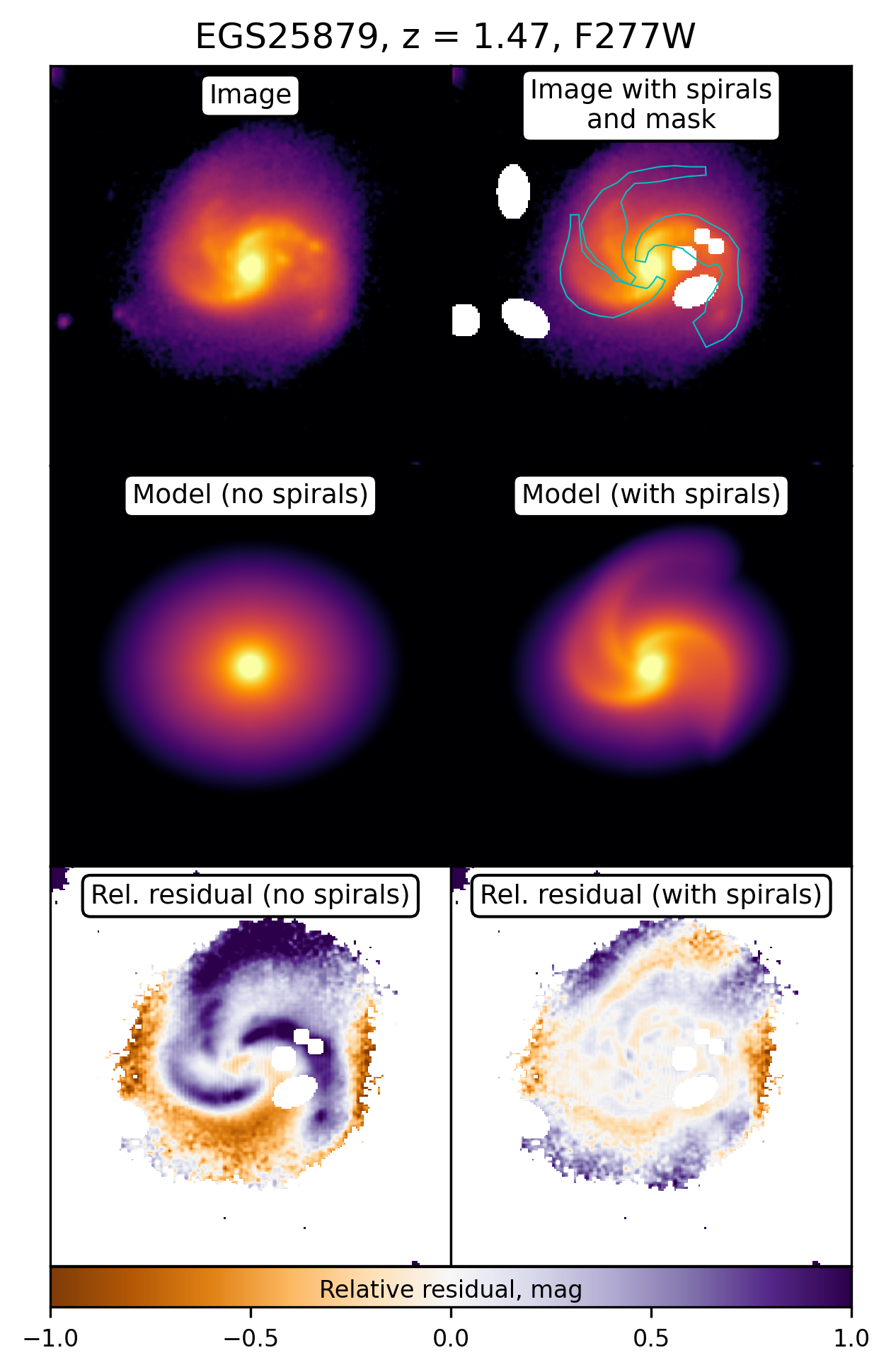}
		\caption{Photometric decomposition model of EGS25879 in the F277W filter shown as an example. From top to bottom: original image, models, and relative residuals (difference between the image and model at each pixel, in magnitudes). The model without spiral arms and the corresponding residuals are presented in the left column, while the model with spiral arms and the corresponding residuals are shown in the right column.}
		\label{fig:EGS25879_F277W}
	\end{figure}

\begin{figure*}[]
		\centering
		\includegraphics[width=0.99\linewidth]{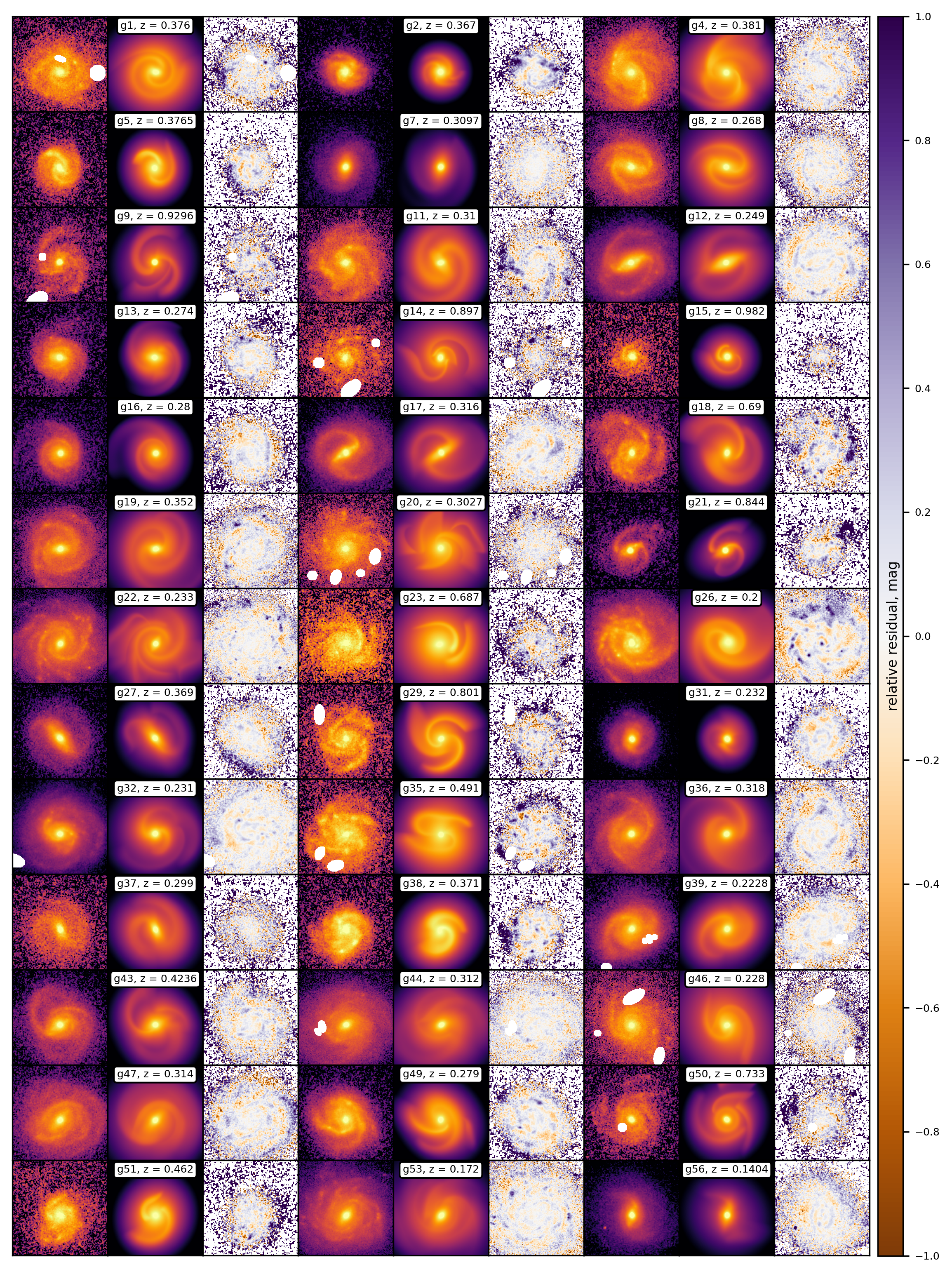}
		\caption{In this mosaic, we display the original images, models, and relative residuals for some galaxies from the COSMOS sample.}
		\label{fig:COSMOS_mosaic_1}
	\end{figure*}

	However, fitting often converges to clearly incorrect results, which can be corrected by using constraints for parameters or by adjusting the initial guess and running the decomposition again, potentially multiple times. As we selected only galaxies with rather regular spiral structure, it limited our sample and made the process of fitting more straightforward. Figure~\ref{fig:EGS25879_F277W} shows an example of the decomposition model for the EGS25879 galaxy in the F277W filter. In Figure~\ref{fig:COSMOS_mosaic_1}, we present a mosaic image showing models of some galaxies from the COSMOS subsample. Mosaic images for all galaxies are available at \url{https://github.com/IVChugunov/Distant_spirals_decomposition}.
	
	\section{Results}
	\label{sec:results}
	With the aid of photometric decomposition, we have measured a large number of galaxy structural parameters. In Table~\ref{tab:results}, we show a portion of our results for a subset of galaxies\footnote{All decomposition results are available at \url{https://github.com/IVChugunov/Distant_spirals_decomposition}.}. Further in this section, we present the results of our fitting.
	
	\begin{table*}[hbt!]
		\begin{threeparttable}
			\caption[]{Some of the structural parameters derived from our decomposition for a subsample of galaxies.}
			\label{tab:results}
			\begin{tabular}{lcccccccccc}
				\toprule
				\headrow Galaxy (filter) & $z$ & $\lambda_\text{rf}$ ($\mu$m) & $M_\text{F814W}$ & $B/T$ & $S/T$ & Disc $h$ (kpc) & $\mu$ (deg) & $l_\psi$ (deg) & $w$ (kpc) & $A_\text{sp}$ \\
				\headrow (1) & (2) & (3) & (4) & (5) & (6) & (7) & (8) & (9) & (10) & (11) \\
				\midrule
				g1 & 0.376 & 0.586 & $-$21.44 & 0.056 & 0.371 & 3.332 $\pm$ 0.049 & 10.8 $\pm$ 3.3 & 415.9 $\pm$ 36.3 & 1.693 $\pm$ 0.248 & 0.257 \\
				g2 & 0.367 & 0.590 & $-$22.13 & 0.390 & 0.110 & 1.550 $\pm$ 0.085 & 23.4 $\pm$ 7.1 & 143.8 $\pm$ 46.9 & 0.785 $\pm$ 0.033 & 0.749 \\
				g4 & 0.381 & 0.584 & $-$22.46 & 0.079 & 0.410 & 3.053 $\pm$ 0.034 & 17.4 $\pm$ 4.1 & 205.3 $\pm$ 44.1 & 2.555 $\pm$ 0.648 & 0.291 \\
				g5 & 0.376 & 0.586 & $-$20.95 & 0.027 & 0.226 & 1.773 $\pm$ 0.013 & 21.2 $\pm$ 3.6 & 185.6 $\pm$ 23.7 & 0.996 $\pm$ 0.052 & 0.452 \\
				\ldots & \ldots & \ldots & \ldots & \ldots & \ldots & \ldots & \ldots & \ldots & \ldots & \ldots \\
				EGS10221 F115W & 1.228 & 0.518 & $-$20.14 & 0.022 & 0.395 & 2.088 $\pm$ 0.069 & 11.3 $\pm$ 4.8 & 93.5 $\pm$ 12.5 & 2.108 $\pm$ 0.008 & 0.541 \\
				EGS10221 F150W & 1.228 & 0.674 & $-$20.14 & 0.028 & 0.362 & 2.207 $\pm$ 0.061 & 17.9 $\pm$ 1.2 & 117.2 $\pm$ 13.3 & 1.855 $\pm$ 0.036 & 0.535 \\
				EGS10221 F200W & 1.228 & 0.893 & $-$20.14 & 0.053 & 0.265 & 3.033 $\pm$ 0.081 & 16.7 $\pm$ 5.6 & 125.7 $\pm$ 11.8 & 1.617 $\pm$ 0.118 & 0.469 \\
				\ldots & \ldots & \ldots & \ldots & \ldots & \ldots & \ldots & \ldots & \ldots & \ldots & \ldots \\
				\bottomrule
			\end{tabular}
			\begin{tablenotes}[hang]
				\item[]Parameters: (1) Name of the galaxy and, possibly, filter. (2) Adopted redshift. (3) Rest-frame wavelength of image. (4) Absolute magnitude in rest-frame F814W filter wavelength. (5) Bulge-to-total luminosity ratio. (6) Spiral-to-total luminosity radio. (7) Disc exponential scale. (8) Average pitch angle of spiral arm. (9) Average azimuthal length of spiral arm. (10) Average width of spiral arm. (11) Asymmetry of spiral structure, see Section~\ref{sec:asymmetry}. The full table, which includes all galaxies and additional parameters, including these for individual spiral arms, is available at \url{https://github.com/IVChugunov/Distant_spirals_decomposition}. For the most parameters of a spiral arm (including pitch angle $\mu$, azimuthal length $l_\psi$, and width $w$) the specified value for a galaxy is an average of the parameters of spiral arms, weighted by their luminosity. Errors for these parameters represent the scatter between individual spiral arms. The error on the disk scalelength $h$ represents a genuine measurement error.
			\end{tablenotes}
		\end{threeparttable}
	\end{table*}
	
\begin{figure}[!ht]
		\centering
		\includegraphics[width=0.99\linewidth]{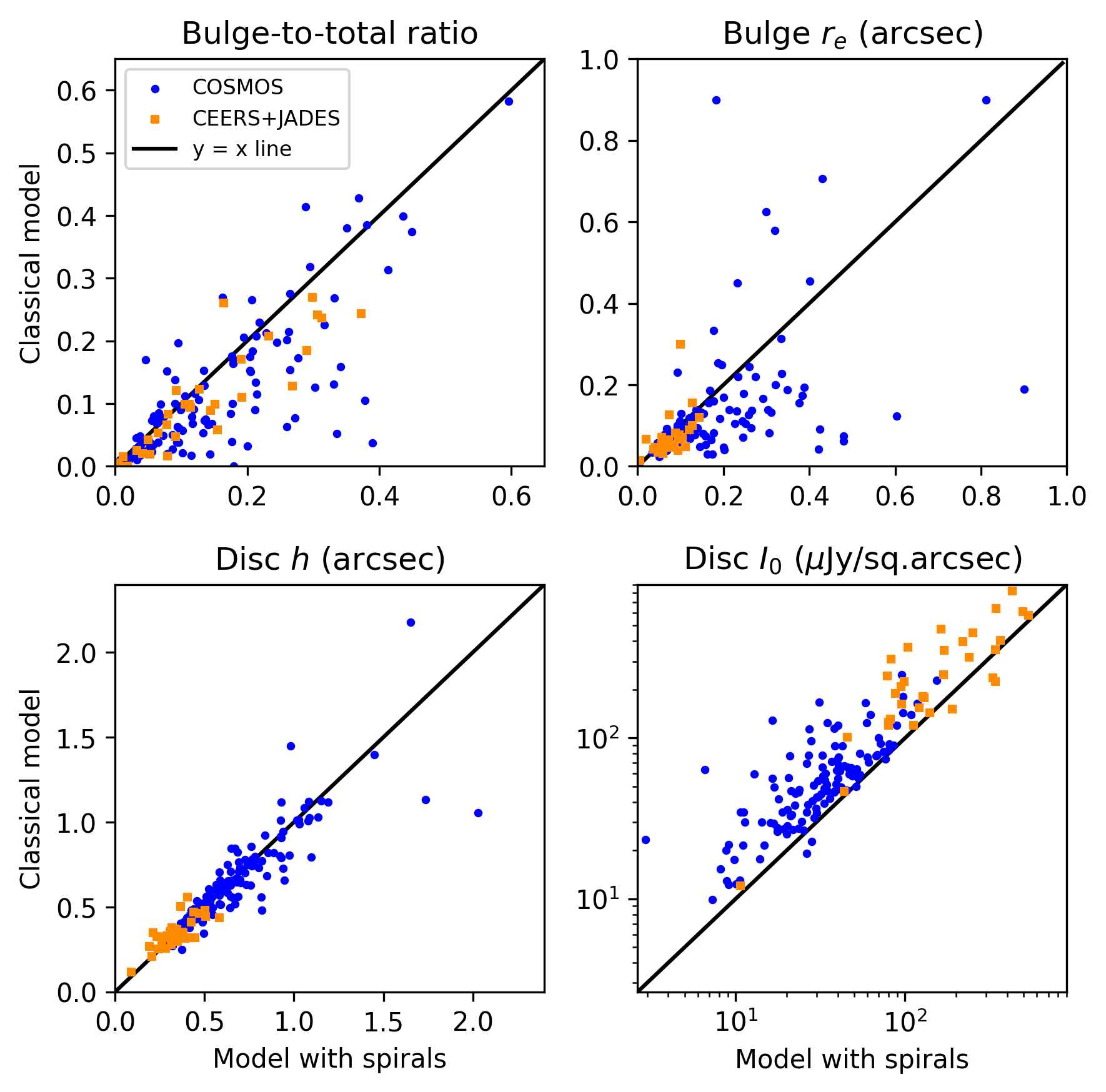}
		\caption{Comparison of some bulge and disc parameters obtained from the decomposition with ``classical'' models. Black solid lines depict a one-to-one ratio.}
		\label{fig:classic_comps}
	\end{figure}

	As our models include ``classical'' components as well as model spiral arms (Section~\ref{sec:decomposition}), we have the opportunity to measure parameters not only for the spiral arms but also for the ``classical'' components. In particular, later we will extensively use the disc exponential scale $h$ as a measure of galaxy size. However, precise measurements of ``classical'' parameters are not a goal of this paper, as we focus primarily on the spiral arm parameters\footnote{In \citelinktext{Chugunov2024}{Paper I} and \citelinktext{Marchuk2024b}{II}, we discussed how the inclusion of spiral arms in a model affects the parameters of other components.}. Nevertheless, in Figure~\ref{fig:classic_comps}, we show some parameters of ``classical'' components and their differences for models with spiral arms and without them. One can see that the bulge-to-total ratio and bulge effective radius increase when spiral arms are added to a model, whereas the disc central surface brightness, as expected, decreases, and the disc exponential scale, on average, does not demonstrate significant deviations from the one-to-one ratio. This is in qualitative accordance with \citelinktext{Chugunov2024}{Paper I} and \citelinktext{Marchuk2024b}{II}, and both papers provide an explanation of such behaviour. In particular, disc central surface brightness drops by 0.55 mag/arcsec$^{-2}$ after the inclusion of spiral arms, and in \citelinktext{Chugunov2024}{Paper I}, this value is 0.5 mag/arcsec$^{-2}$. \citet{Mosenkov2024} reports that the central disc surface brightness drops by 0.4--0.8 mag/arcsec$^{-2}$ if the spiral arms are masked. Bulge-to-total ratio increases by a factor of 76\% after the inclusion of spiral arms to model and effective radius increases by 65\%, whereas in \citelinktext{Chugunov2024}{Paper I} there are significantly smaller values of 33\% and 20\%, respectively. This discrepancy could be attributed to a number of reasons, out of the most obvious being that photometric properties of bulges are different for our dataset compared to \citelinktext{Chugunov2024}{Paper I} because of different rest-frame wavelength, smaller redshift and higher luminosity of galaxies. Nevertheless, the general trend remains the same and the explanation from \citelinktext{Chugunov2024}{Paper I} can be applied to these results as well.
	
	In the following subsections, we will first focus on the general parameters of spiral arms and on their simple dependence on redshift $z$ or lookback time $t_L$ (Section~\ref{sec:general_results}). After that, we discuss the dependence of the observed parameters on rest-frame wavelength $\lambda_\text{rf}$ (Sections~\ref{sec:band-shifting} and~\ref{sec:discerning}). There is also a number of parameters that we have measured but they are less important and do not show any remarkable variations with redshift or wavelength, but they raise questions about the exact form of the distribution of light in spiral arms and will be discussed in the following work (Chugunov \& Marchuk, in prep.).
	
	We note that for most parameters that we dealing with in the following, their values can be attributed to each spiral arm individually and differ significantly between arms in a galaxy, e.g. as in the case of pitch angles. In most such cases, we attribute a single value of the parameter to a galaxy as an average among its spiral arms weighted by their luminosity. Among all the parameters that will appear in this Section, the only exceptions are spiral-to-total luminosity ratio, which is a sum of individual spiral arm contributions to the total luminosity (Section~\ref{sec:S-T}), the spiral arms extent, which is instead a maximum value among spiral arms (Section~\ref{sec:extent}), and the asymmetry which is calculated in another way, described in the corresponding Section~\ref{sec:asymmetry}.
	
	\subsection{General parameters of spiral arms}
	\label{sec:general_results}
	Regarding the simple dependence of the parameters on $z$ or $t_L$, the most notable findings relate to the overall shape of spiral arms, specifically the pitch angle and azimuthal length.
	
	\subsubsection{Pitch angles}
	\label{sec:pitch}
	
\begin{figure}[!ht]
		\centering
		\includegraphics[width=0.99\linewidth]{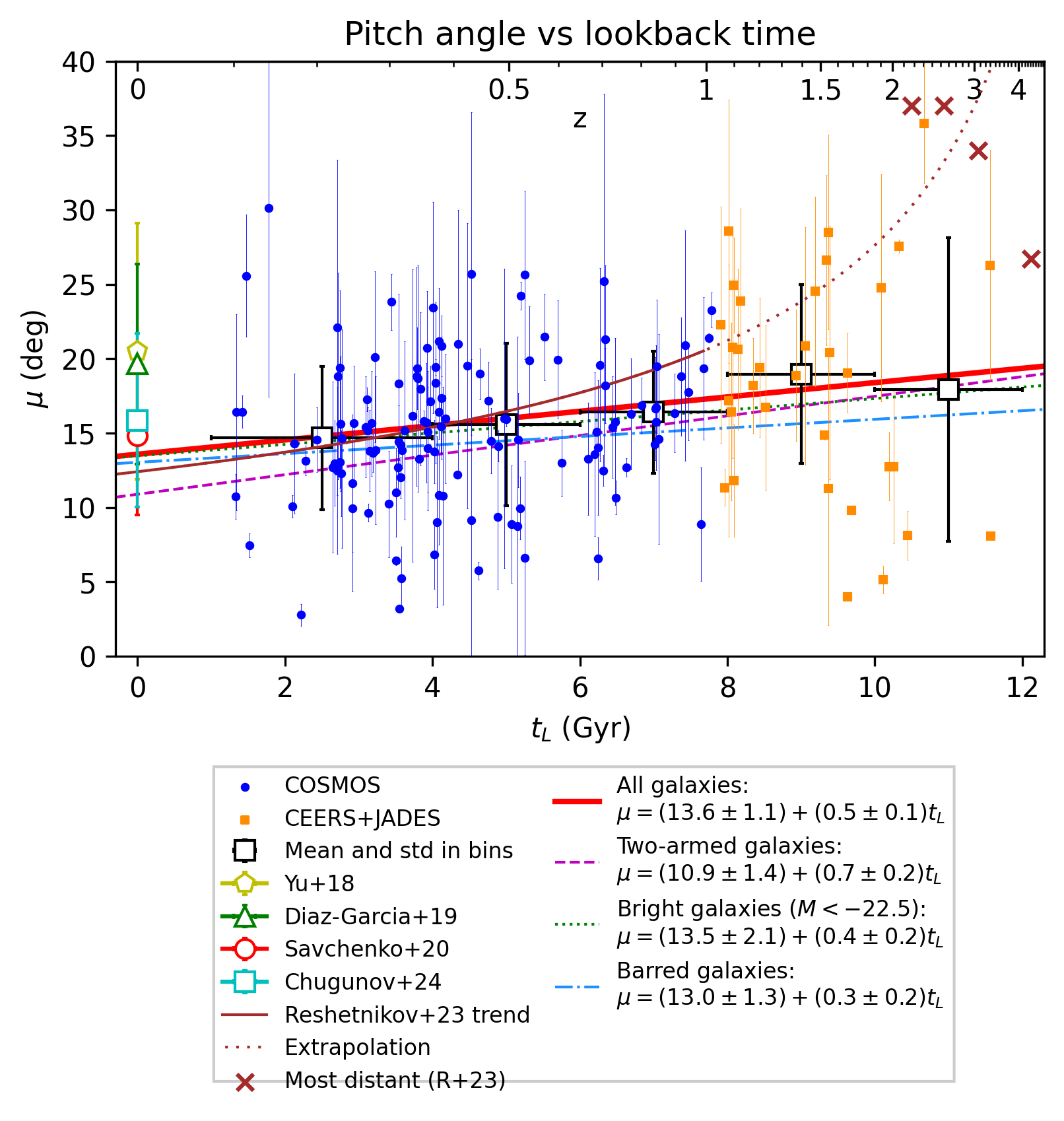}
		\caption{Pitch angle dependence on lookback time $t_L$. In this and subsequent illustrations, black squares with bars represent average values in $t_L$ bins: horizontal bars show the binning range and vertical bars show the standard deviation. Each point corresponds to an averaged pitch angle for a galaxy. Blue circular points represent galaxies from COSMOS, and orange squares represent galaxies from CEERS and JADES. Pitch angles for the CEERS and JADES subsamples are shown as measured in only one band, namely the F200W or F210W filter, whichever is available, to avoid clutter (subsequent diagrams against lookback time have the same feature). Various lines represent linear regressions for the entire sample of galaxies and for a few subsamples: bright galaxies, two-armed galaxies, and barred galaxies, as specified in the legend. Measurements from~\citet{Yu2018a}, \citet{Diaz-Garcia2019}, \citet{Savchenko2020}, and \citet{Chugunov2024} are also shown, as well as a general linear trend with $z$ from~\citet{Reshetnikov2023}. Individual measurements of pitch angles for the most distant known spiral galaxies, as reported by~\citet{Law2012, Yuan2017, Wu2023, Tsukui2021} and compiled in~\citet{Reshetnikov2023}, are also presented.}
		\label{fig:mu-time}
	\end{figure}

	Firstly, we examine the pitch angles $\mu$ of galaxies. As we adopt a model that produces spiral arms with varying pitch angles, some simplification is necessary. To assign a single pitch angle value to an arm, we perform a linear fit to the $\log r(\psi)$ function (defined by the best-fitting parameters; see Equations~\ref{eq:r_psi} and~\ref{eq:k_m}). This approach effectively fits a logarithmic spiral to the analytical function that best describes the shape of the spiral arm. Subsequently, as with most other parameters, we compute the luminosity-weighted average of pitch angles of the individual arms to determine an averaged pitch angle value for the galaxy. The average value of $\mu$ over the sample is $16.0$ degrees with a standard deviation of $5.7$ degrees. The dependence between $\mu$ and lookback time $t_L$ is depicted in Figure~\ref{fig:mu-time}. The pitch angle gradually increases with $t_L$ at a rate of 0.5 degrees per Gigayear. Though such an increase is within the errors for individual galaxies, it implies a decrease as we move closer to the modern epoch. A similar, albeit slightly stronger trend has been found in previous studies \citep{Reshetnikov2022, Reshetnikov2023}. In this paper, we confirm these results for an intersecting sample but use a different method for measuring the pitch angle. The subsample of galaxies from CEERS and JADES suggests that this trend continues up to $t_L \approx 11$ Gyr, or $z \approx 2.5$. \citet{Reshetnikov2023} also mention that the known pitch angles for four very distant galaxies ($2 < z < 4.5$) are indeed larger than the average values over the range $0 \leq z \leq 1$. At the same time, \citet{Davis2012} did not find any dependence of the pitch angle on redshift, which is possibly because the Fourier-based method used was unable to properly detect pitch angle variations \citep{Yu2018a}.
	
	Note that the average pitch angles of galaxies in this work, as well as the average values in~\citet{Savchenko2020} and \citelinktext{Chugunov2024}{Paper I}, are noticeably lower than the values from~\citet{Yu2018a} and~\citet{Diaz-Garcia2019}. The most plausible explanation for this is that~\citet{Yu2018a} and~\citet{Diaz-Garcia2019} used Fourier-based techniques and were able to measure pitch angles in spiral galaxies of {\it all} types, whereas our work and~\citet{Savchenko2020} employ methods that allow measurements only in galaxies with distinct and recognisable spiral arms. Therefore, samples of galaxies whose parameters can be measured by slices or photometric decomposition are biased towards multi-armed and grand-design types, which correspond to earlier Hubble types and smaller pitch angles.
	
\begin{figure}[!ht]
		\centering
		\includegraphics[width=0.99\linewidth]{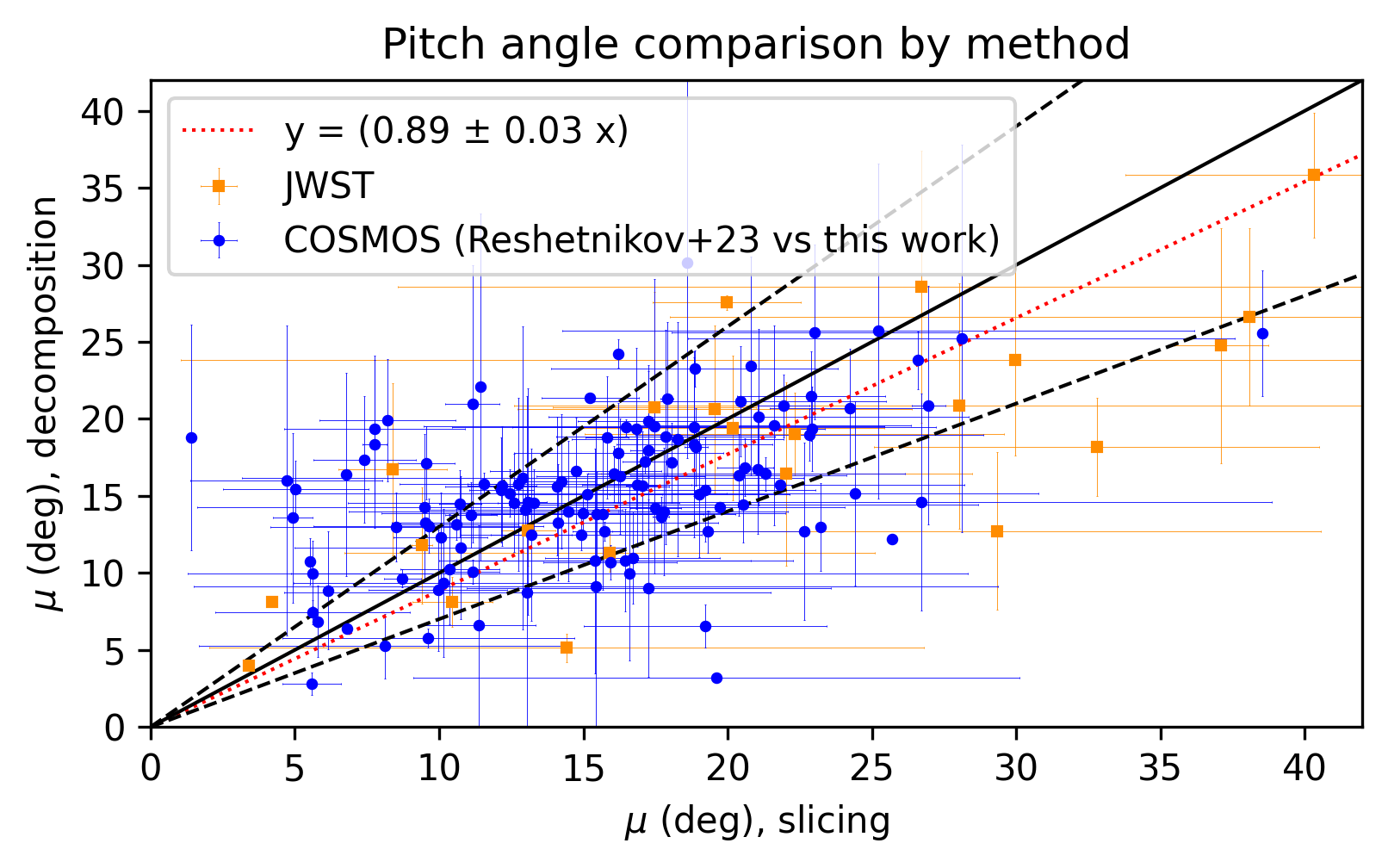}
		\caption{Comparison of average pitch angles obtained using the slicing method and decomposition. For the COSMOS subsample, pitch angle measurements with the slicing method are taken from~\citet{Reshetnikov2023}, whereas, for the JWST data, the slicing method was applied in this work. The solid line depicts a 1:1 relation, while the dashed lines show a 30\% error. The red dotted line depicts a linear approximation.}
		\label{fig:comp_Resh}
	\end{figure}

	This trend becomes more pronounced if we consider only two-armed galaxies (0.7 deg/Gyr) and remains positive for the brightest galaxies ($M_\text{F814W} < -22.5$) and for barred ones. As our sample largely intersects with that from~\citet{Reshetnikov2023} where pitch angles were measured using the slicing method, we can compare the measured pitch angles for individual objects; this comparison is shown in Figure~\ref{fig:comp_Resh}. Also, for galaxies from the JWST subsample, we measured pitch angles using a similar slicing method and included them in this comparison.
	
	The consistency seems to be not very good; however, there are a number of possible reasons. Needless to say, the methods of measuring the pitch angle differ: when the slicing method measures the pitch angle over only a part of the spiral arm, it can easily differ from the value measured by decomposition, as the pitch angle often varies significantly along the spiral arm. Additionally, in many cases, even the number of spiral arms for an individual galaxy is different from~\citet{Reshetnikov2023}, which, combined with the large differences in pitch angle between individual arms, can strongly affect the measured pitch angles (notice the large error bars in Figure~\ref{fig:comp_Resh}, representing the scatter between spiral arms). We also note that in the mentioned work, the inclination of galaxies was not taken into account, although it is not very large in any case. It is worth noting that, in the domain of higher pitch angles, measurements obtained using the slicing method are slightly higher than those from decomposition, and the linear fit coefficient is less than unity. This discrepancy can likely be attributed to the slicing method's difficulty in identifying the brightness peak of the spiral arm near the galactic centre, where the brightness gradient is steep. As a result, the peak position is estimated to be too close to the galactic centre, causing the measured $r(\psi)$ to appear overly steep. This may also explain why~\citet{Reshetnikov2023} observed a steeper trend of $\mu$ vs. $z$ compared to the results of this work.
	
	In particular, let us consider one of the most outlying galaxies, g35 (see Figure~\ref{fig:COSMOS_mosaic_1}), with $\mu = 16.0$ degrees in this study and $\mu = 4.7$ degrees in~\citet{Reshetnikov2023}. The shape of its spiral arms is highly non-logarithmic, as pitch angles reach very high values near the centre of the galaxy and become close to zero or even negative in the outer parts. In this specific galaxy,~\citet{Reshetnikov2023} examined the shape of the spiral arms only in the most prominent middle part of their length, where the pitch angle is indeed close to zero. Meanwhile, our model of these arms traces them along all their extent, capturing parts with high pitch angle, yielding a much higher average value. To summarise, different methods of measuring pitch angles often yield different results for the same galaxies --- see, for example, Figure~10 in~\citet{Savchenko2020} or Figure~8 in \citet{Yu2019}.
	
	\subsubsection{Pringle--Dobbs test}
	\label{sec:Pringle-Dobbs}
	Having measured the pitch angles of spiral arms, we can apply the Pringle–Dobbs test~\citep{Pringle2019} to investigate possible mechanisms of spiral structure formation. The main idea behind the test is as follows: if the spiral arm has a transient or tidal nature, its pitch angle $\mu$ decreases with time $t$ according to $\cot \mu \propto t$. Consequently, if the spiral arms in a sample of galaxies are predominantly transient or tidal, we would expect the value of $\cot \mu$ to be uniformly distributed. This implies that the `age' of the spiral arms is random and also uniformly distributed.
	
\begin{figure}[!ht]
		\centering
		\includegraphics[width=0.99\linewidth]{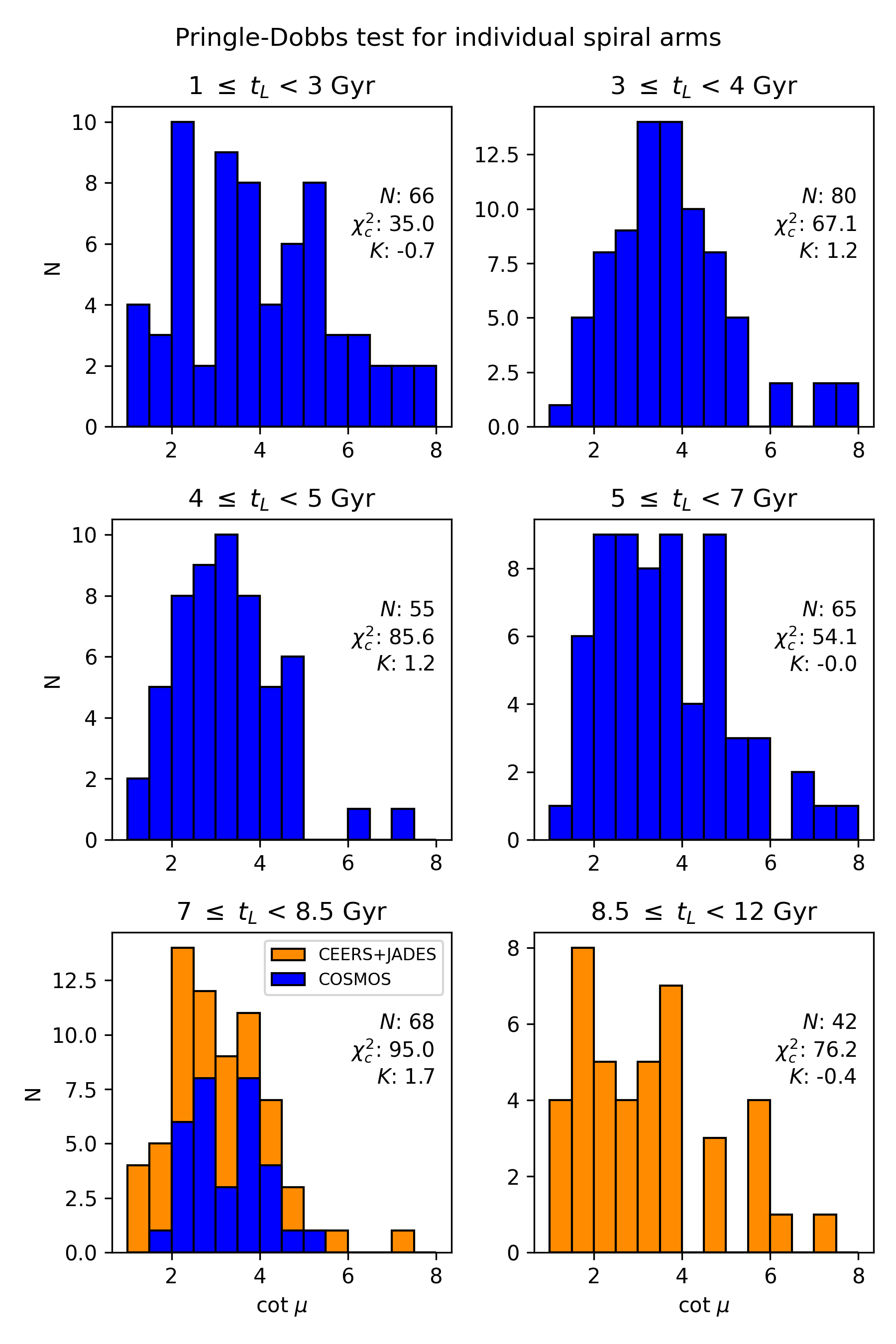}
		\caption{Application of the Pringle--Dobbs test to our sample of measured individual spiral arms binned by lookback time. Bins are chosen so that there are roughly the same number of spiral arms in each bin. For each bin, the number of spiral arms $N$, the chi-square statistic for uniform distribution $\chi^2_c$ (corrected for the unequal number of arms), and the kurtosis $K$ are specified.}
		\label{fig:Pringle-Dobbs}
	\end{figure}

	We divide our galaxies into bins based on lookback time $t_L$ and perform the Pringle–Dobbs test for each bin. The resulting distributions are shown in Figure~\ref{fig:Pringle-Dobbs}. Both visually and using statistical indicators (namely, the chi-square statistic for uniform distribution and kurtosis), it is evident that $\cot \mu$ is distributed more uniformly in the first (1–3 Gyr) bin, and the last (8.5–12 Gyr) bin also shows a higher degree of uniformity than most of the intermediate bins. Studies of local galaxies indicate that $\cot \mu$ is distributed nearly uniformly over some range of $\mu$ \citep{Pringle2019, Lingard2021}. For distant galaxies, \citet{Reshetnikov2022, Reshetnikov2023} found that the distribution is more or less uniform for smaller $t_L$ but becomes increasingly non-uniform closer to $z \sim 1$ ($t_L \sim 8$ Gyr). Our result is consistent with theirs, but the observed return to a more uniform distribution at $t_L > 8.5$ Gyr is a new finding.
	
	Analysing the $\cot \mu$ distribution by individual arms maximises the number of data points, but it becomes more difficult to determine the uniformity of distributions when using pitch angles of individual galaxies or when considering subsamples (such as bright, two-armed, or barred galaxies). Nevertheless, the degree of non-uniformity in the $\cot \mu$ distributions in these cases is consistent with what is observed in Figure~\ref{fig:Pringle-Dobbs}.
	
	This behaviour (specifically, that $\cot \mu$ is the least uniform at intermediate $t_L$ and the most uniform at the smallest and largest $t_L$) requires further investigation. If this pattern holds true, we propose the following simplistic explanation. It is well known that galactic interactions were much more common in the early Universe than they are today~\citep{Conselice2007}, and these interactions are capable of producing tidal arms. Theoretically, they can also induce density waves in discs, which may also contribute to the emergence of spiral structure~\citep{Dobbs2014}.
	
	We speculate that at $z > 1$, spiral structures are predominantly tidal in nature, which aligns with the uniform $\cot \mu$ distribution and higher $\mu$ values on average. Later, tidal spiral arms mostly disappear, but tidally-induced density waves remain for a few Gyr, so most of the observed spiral arms in this period are density waves, resulting in a highly non-uniform $\cot \mu$ distribution. Finally, as we approach the modern epoch, density waves begin to fade, and transient spiral structures become the most common, leading to a more uniform $\cot \mu$ distribution once again.
	
	\subsubsection{Pitch angle variations}
	\label{sec:variations}
\begin{figure}[!ht]
		\centering
		\includegraphics[width=0.99\linewidth]{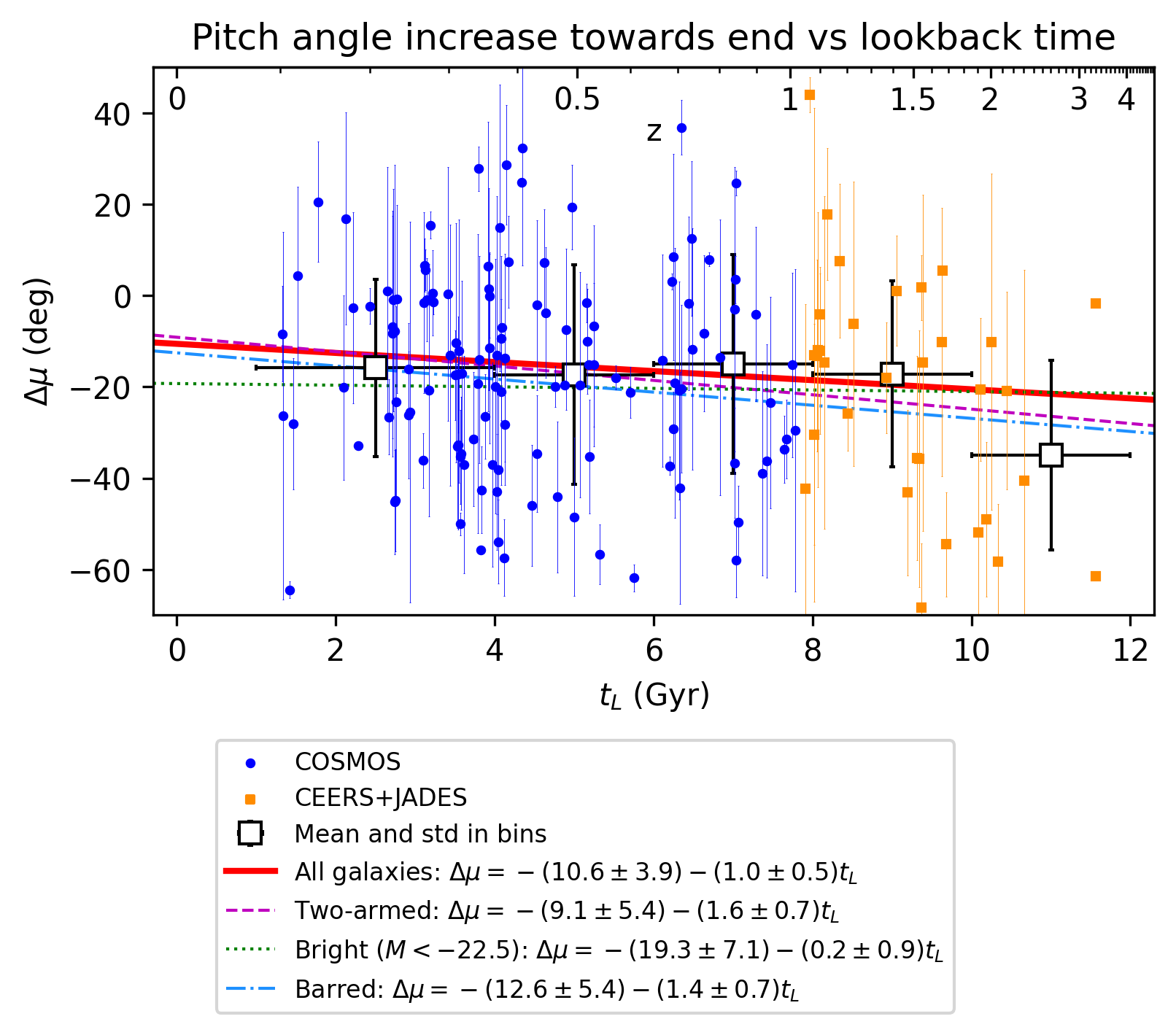}
		\caption{Same as Figure~\ref{fig:mu-time}, but for the pitch angle increase towards the end of the arm $\Delta \mu$ instead of the pitch angle $\mu$.}
		\label{fig:dmu-time}
	\end{figure}

	We measure how the pitch angle of a spiral arm increases or decreases from the beginning (inner parts of a galaxy) towards the end (outer parts). Again, similarly to the case of the average pitch angle (Section~\ref{sec:pitch}), our model is capable of producing spiral arms with more complex pitch angle behaviour, beyond a simple monotonic increase or decrease. Therefore, we fit $r(\psi)$ obtained from decomposition with a function where $\mu$ changes linearly with $\psi$ (which is essentially Equation~\ref{eq:r_psi} but with $n$ ranging from 1 to 2) and consider the difference in $\mu$ from the beginning to the end of this fit, referring to this parameter as $\Delta \mu$. In Figure~\ref{fig:dmu-time}, we show $\Delta \mu$ displayed against $t_L$.
	
	First, we find that, on average, pitch angles decrease by $-17.3$ degrees from the beginning to the end of a spiral arm. In our sample, 127 of 159 galaxies, or 80\%, have decreasing pitch angles (more tightly wound arms on the disc's periphery). If we consider only two-armed or bright galaxies, the proportions are 81\% and 84\%, respectively. This is higher than the 64\% observed for local two-armed galaxies found by \citet{Savchenko2013}, but in any case, it indicates that pitch angles tend to decrease along the spiral arm. Our result, indicating highly variable pitch angles in spirals, also goes in line with the work of~\citet{Savchenko2020}, who concluded that pitch angle can vary by 56\% of the mean value. On another side,~\citet{Mosenkov2024} compared pitch angles inside and outside the $R_{25}$ and found that their values can differ significantly, but galaxies with increasing and decreasing pitch angles are distributed equally. However, in most cases, we do not trace spiral arms beyond $R_{25}$, and this result does not contradict ours.
	
	Second, we note that $\Delta \mu$ changes with time in a non-uniform way. Throughout most of the $t_L$ range we consider, namely up to 9 Gyr, the average $\Delta \mu$ remains nearly constant and moderate, with an average value of $-15.6$ degrees for $t_L < 9$ Gyr. After this point, $\Delta \mu$ decreases abruptly, and at $t_L \geq 9$ Gyr, the average $\Delta \mu$ is $-29.9$ degrees, with 16 out of 19 galaxies showing a negative value. Even when considered relative to the average $\mu$, $\Delta \mu / \mu$ has a larger absolute value at $t_L > 9$ Gyr than at later times. Although $\mu$ is larger at early epochs, its variation with time is less than a factor of two.
	
	\subsubsection{Azimuthal lengths}
	\label{sec:length}
\begin{figure*}[!ht]
		\centering
		\includegraphics[width=0.99\linewidth]{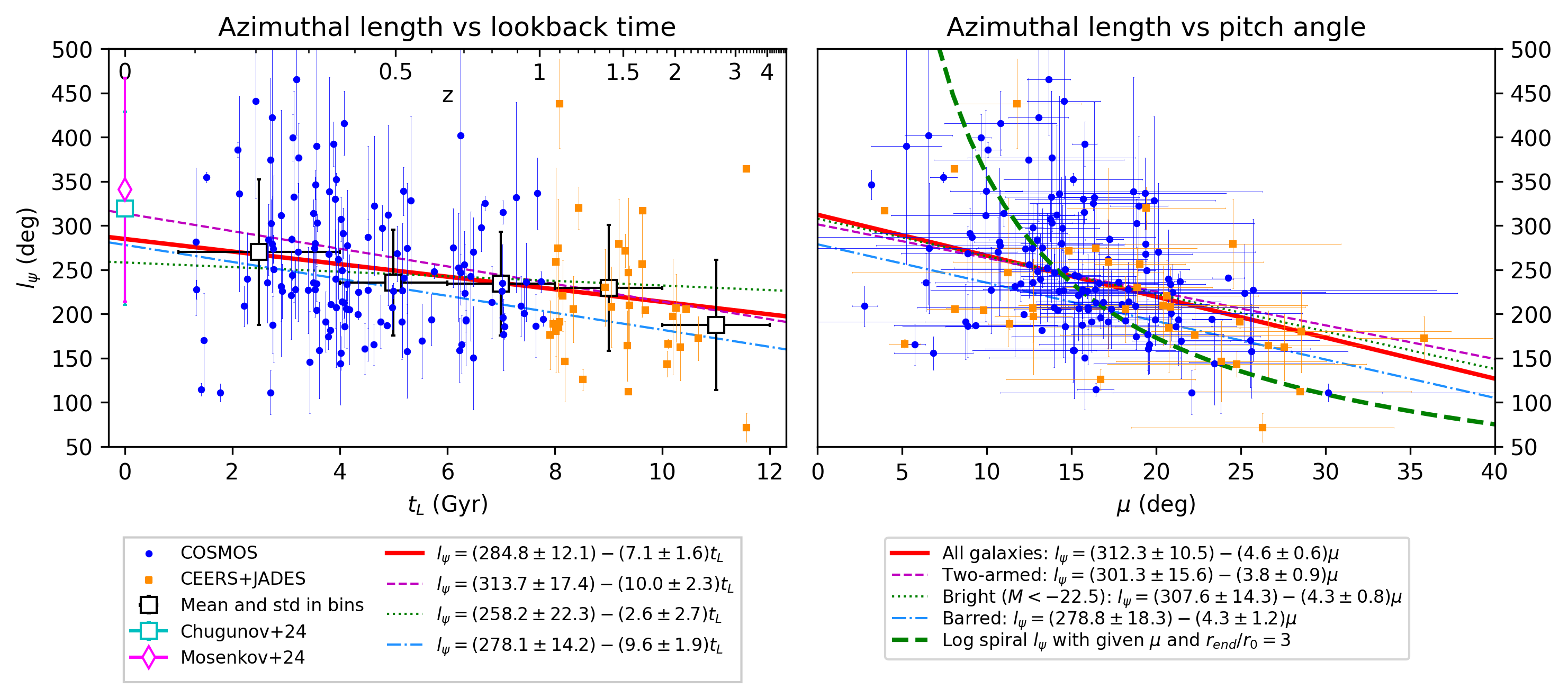}
		\caption{Left: same as Figure~\ref{fig:mu-time}, but for azimuthal length $l_\psi$ instead of pitch angle $\mu$. Measurements from~\citet{Chugunov2024, Mosenkov2024} are also shown. Right: azimuthal length $l_\psi$ versus pitch angle $\mu$ diagram. The curve represents $l_\psi$ of a logarithmic spiral with a given $\mu$ that has ending radius $r_\text{end} = 3\times r_0$, where $r_0$ is the beginning radius of the arm.}
		\label{fig:length-time}
	\end{figure*}

	Another parameter we consider is the azimuthal length $l_\psi$, which is defined for an individual spiral arm as the azimuthal angle between its beginning and end. In Figure~\ref{fig:length-time}, we show $l_\psi$ displayed against $t_L$. It can be observed that $l_\psi$ decreases significantly with increasing $t_L$ at a rate of $-7$ degrees per Gyr, which implies that spiral arms become longer as we approach the modern epoch. This is also consistent with our measurements of length from the \citelinktext{Chugunov2024}{Paper I} data for local galaxies, which are larger, on average, than for any meaningful part of this sample.
	
	The decrease of $l_\psi$ with $t_L$ is observed regardless of the subsample considered. For barred galaxies, the rate of decrease is $-10$ degrees per Gyr, and it is the same for two-armed galaxies. Notably, $l_\psi$ tends to be smaller in barred galaxies than in unbarred ones by about 30 degrees, on average, at the same $t_L$ or the same $\mu$. This may be because spiral arms in barred galaxies usually start from the ends of the bar, positioning their beginning farther from the centre than in unbarred galaxies. Assuming similar pitch angles, one might expect that spiral arms in barred galaxies reach the radius where the spiral structure disappears at a smaller $l_\psi$. When considering only bright galaxies, the measured rate of $l_\psi$ decrease is $-3$ degrees per Gyr.
	
	In principle, such a trend could simply be an observational effect. If a spiral galaxy is contaminated by noise in an image and is poorly resolved, it is likely more difficult to trace spiral arms to their full extent. Specifically, the observed extent of spiral arms (Section~\ref{sec:extent}) is, on average, 0.67 of the optical radius $R_{25}$ in the rest-frame whereas~\citet{Savchenko2020} and~\citet{Mosenkov2024} found that spiral arms in local galaxies can be traced up to $R_{25}$. Indeed, more distant galaxies have the poorer spatial resolution, at least up to $z \approx 1.7$ \citep{Melia2018}, and lower apparent surface brightness due to cosmological surface brightness dimming \citep{Calvi2014}. To determine whether these effects significantly impact the measurement of $l_\psi$, we performed a validation of our results as described in Section~\ref{sec:validation}, and we conclude that at least part of the observed trend is evolutionary rather than observational.
	
	Note that a simultaneous decrease of azimuthal length and increase of pitch angle (see the right panel in Figure~\ref{fig:length-time}) is expected. If a spiral arm has high pitch angle, the radius of a point on the arm grows fast with azimuthal angle, and the high azimuthal length of the arm in this case would imply a very large ratio between the beginning and ending radii $r_0$ and $r_\text{end}$. Considering that spiral arms usually extend to a limited part of a disc, neither in the very centre nor in the far periphery, we expect the ratio $r_\text{end} / r_0$ to be limited.
	
	\subsection{Band-shifting effects}
	\label{sec:band-shifting}
	When galaxies with different $z$ are observed in a fixed band, the rest-frame wavelength of the observed light varies. This can distort results concerning the evolution of separate components, as demonstrated for bars in~\citet{Menendez-Delmestre2024}. To address this issue, we analyse multi-band data for CEERS and JADES images to distinguish band-shifting effects from genuine evolutionary changes in galaxies. 
	
\begin{figure}[!ht]
		\centering
		\includegraphics[width=0.99\linewidth]{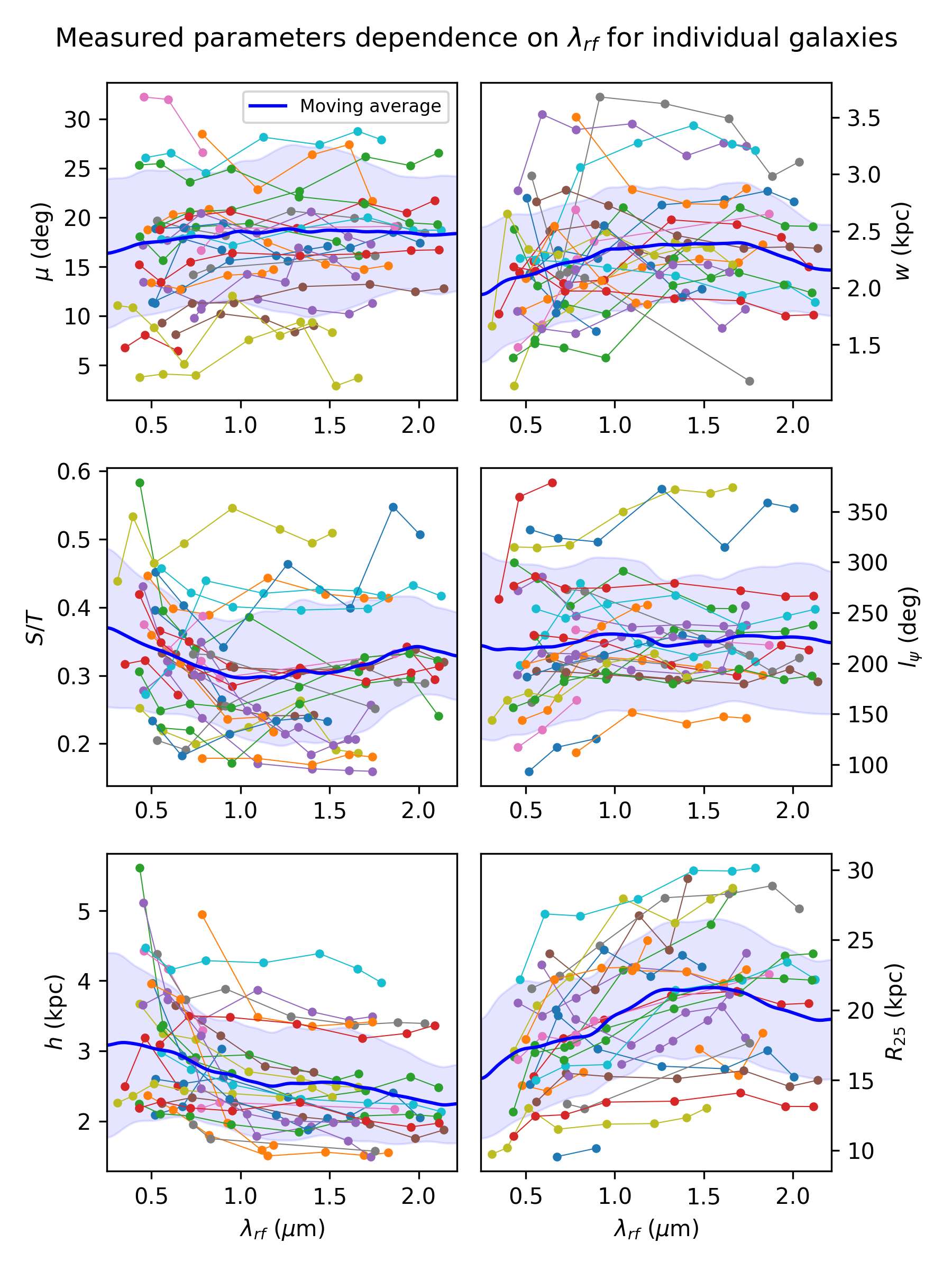}
		\caption{For individual galaxies from the CEERS and JADES surveys, the measured values of six parameters are displayed against the rest-frame wavelength $\lambda_\text{rf}$. Each dot represents a parameter measured for an individual galaxy at a single wavelength. Measurements for a single galaxy at different wavelengths are represented by dots of the same colour and are connected by a line. The thicker blue line shows the smoothed moving average with a window of 0.2~$\mu$m half-width.}
		\label{fig:params-lambda}
	\end{figure}

	To gain a qualitative understanding of which parameters may be distorted by band-shifting effects, we can compare parameters measured at different wavelengths for images of a single galaxy, which is feasible for objects from the CEERS and JADES surveys. In Figure~\ref{fig:params-lambda}, the measured values of a few parameters are charted against the rest-frame wavelength $\lambda_\text{rf}$, along with their moving averages.
	
	Indeed, there is a systematic change with $\lambda_\text{rf}$ for some parameters, such as the spiral arm width $w$ and the contribution of the spiral arms to the total luminosity of a galaxy (spiral-to-total luminosity ratio, or $S/T$ for short). Other parameters, such as pitch angle $\mu$ or azimuthal length $l_\psi$, both discussed in Section~\ref{sec:pitch}, mostly remain unchanged across different rest-frame wavelengths.
	
	For some parameters like $w$, it makes more sense to analyse them relative to the size of a galaxy, as they are clearly connected (see, e.g., our \citelinktext{Chugunov2024}{Paper I}). As a measure of galaxy size, we adopt the disc exponential scale $h$, which turns out to be smaller at longer wavelengths. Conversely, $R_{25}$ (here, it is the radius of the 25 mag/arcsec$^{-2}$ isophote for a {\it given} wavelength) increases at longer wavelengths, likely because the typical SED of a galaxy has a positive slope in this part of the spectrum (see, for example, figure 4 in~\citealp{Nersesian2020}). We note that in our \citelinktext{Marchuk2024b}{Paper II}, we found that $w$, $h$, and $S/T$ all indeed vary strongly with wavelength.
	
	Interestingly, $w$, $h$, and $S/T$ all demonstrate stronger variations with wavelength at smaller $\lambda_\text{rf}$, but this variation becomes weaker or even reverses beyond 1.0--1.5~$\mu$m. We observed this behaviour in our \citelinktext{Marchuk2024b}{Paper II} over the same wavelength range from M~51. A similar effect (but for other parameters) was found in~\citet{Ren2024}, who analysed the dependence of various morphological indicators on rest-frame wavelength and found these effects to be negligible at $\lambda_\text{rf} > 1~\mu$m.
	
	We can explain both of these effects in the following way. First, the overall variation of $w$ and $S/T$ with $\lambda_\text{rf}$ can be explained as presented in our \citelinktext{Marchuk2024b}{Paper II}. Star formation in galaxies is strongly concentrated in spiral arms, and so are the youngest stars, which emit in bluer bands. Consequently, spiral arms appear most prominent in bluer bands and have bluer colours than other parts of the galaxy. Additionally, the youngest stars are closely tied to their birth locations and cannot propagate far from the site of ongoing star formation, making spiral arms outlined by the youngest stars relatively narrow. In contrast, older stars have enough time to migrate from their formation sites, and their radiation, primarily observed at longer wavelengths, makes the apparent width of the arm greater. This interpretation is supported by the results of~\citet{Pessa2023}, who found that spiral arms become wider when older stellar populations are considered (see also the resolved stellar populations for IC\,342 obtained by Euclid in \citealp{Euclid}). The variation of $h$ with $\lambda_\text{rf}$ is also consistent with \citelinktext{Marchuk2024b}{Paper II} and with \citet{Casasola2017} over the relevant range of $\lambda_\text{rf}$. For further discussion of this effect, we refer the reader to the latter work.
	
	Expanding on this idea, it is possible to explain why neither $S/T$ nor $w$ change significantly at $\lambda_\text{rf} > 1.0$--$1.5~\mu$m. The ratio between the contributions of old and young stellar populations increases with wavelength in the optical range, peaks at 1.5–3 $\mu$m, and remains relatively constant near these wavelengths (see, for example, figure 4 in~\citealp{Nersesian2020}). Therefore, a slight change in wavelength at $\lambda_\text{rf} < 1 \mu$m represents a considerable change in the highlighted population, leading to variations in $S/T$ and $w$. However, near 1.5–3 $\mu$m, any wavelength highlights a similar stellar population, resulting in consistent values of $S/T$ and $w$.
	
	\subsection{Discerning band-shifting and evolutionary effects}
	\label{sec:discerning}
	We now analyse the dependence of various parameters on $t_L$ and rest-frame wavelength $\lambda_\text{rf}$ in a more quantitative manner. For each parameter, we fit it using a bilinear function of redshift and rest-frame wavelength, allowing us to separately assess the dependencies on these parameters. Only galaxies with $\lambda_\text{rf} < 1.5 \mu$m are considered, for reasons discussed in Section~\ref{sec:band-shifting}. As demonstrated in the same section, some parameters do not vary significantly with $\lambda_\text{rf}$, making this form of analysis unnecessary, as the observed dependence on $t_L$ is not affected by band-shifting effects. In fact, all parameters previously discussed in Section~\ref{sec:general_results} are independent of $\lambda_\text{rf}$, and thus, we do not consider them further in this section.
	
	\subsubsection{Spiral-to-total luminosity ratio}
	\label{sec:S-T}
\begin{figure}[!ht]
		\centering
		\includegraphics[width=0.99\linewidth]{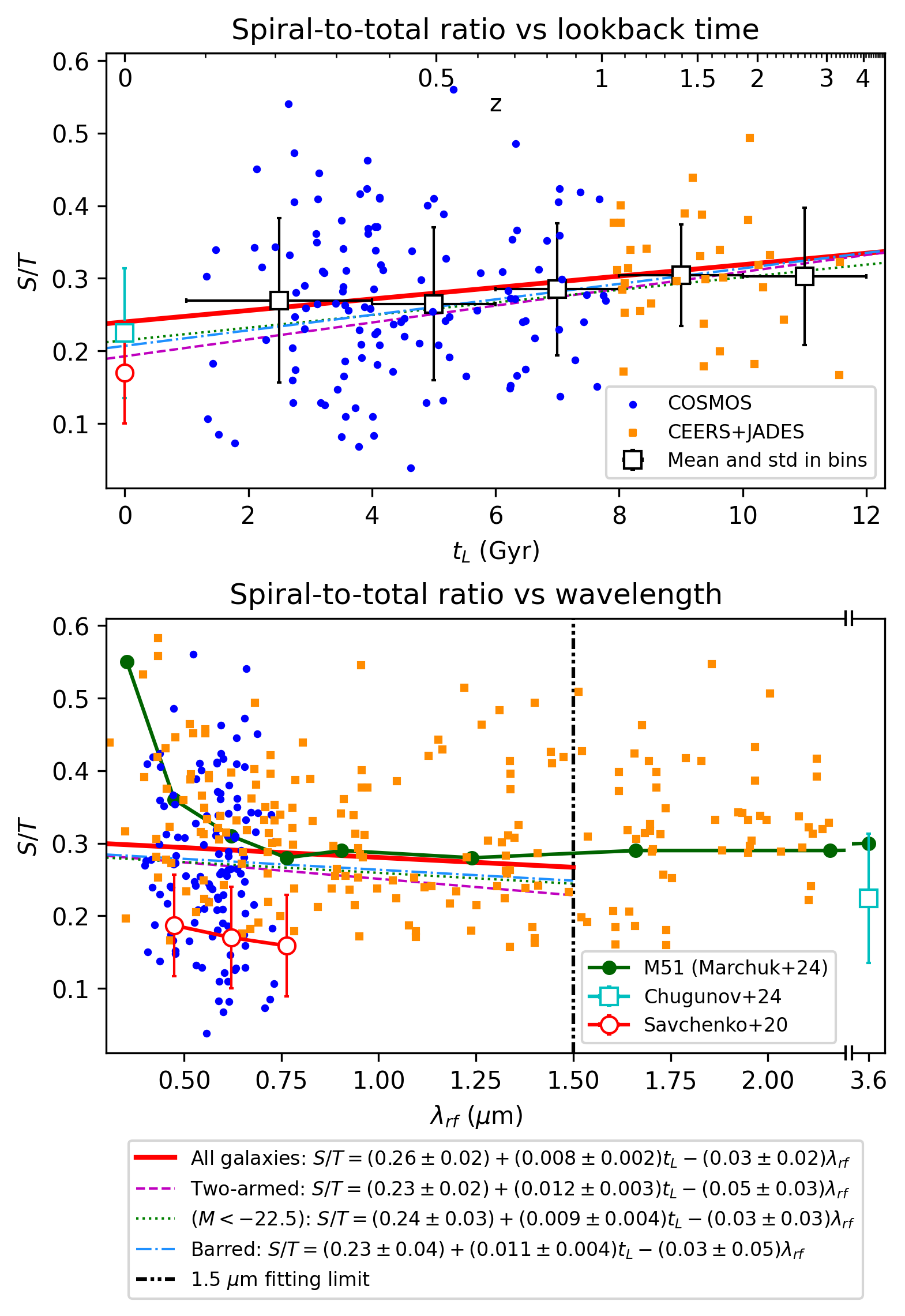}
		\caption{Dependence of spiral-to-total luminosity ratio on lookback time (top) and on rest-frame wavelength (bottom). Different lines represent the projection of the bilinear regression (as specified in the legend) for all sample and different subsamples. Measurements from~\citet{Marchuk2024b, Chugunov2024, Savchenko2020} are also shown.}
		\label{fig:ST-time-lambda}
	\end{figure}

	First, we consider the spiral-to-total luminosity ratio $S/T$ (see Figure~\ref{fig:ST-time-lambda}). Note that this parameter represents the sum of individual spiral arm fractions, rather than a weighted average as used for most other parameters. This approach is more appropriate for measuring the fraction of a galaxy's luminosity attributed to the entire spiral pattern. We found that the average $S/T$ value over the sample is 0.28, with a standard deviation of 0.10. There is an observed increase with $t_L$ at a rate of 0.08 per 10 Gyr and a decrease with increasing $\lambda_\text{rf}$, at a rate of $-0.03$ per 1 $\mu$m.
	
	Both trends vary depending on which subsample is considered (e.g., two-armed, bright, or barred galaxies), but they remain qualitatively similar. The dependence on $t_L$ suggests that the spiral arm fraction decreases with time; however, this is likely a selection effect, as it can be challenging to identify distant galaxies as spirals if they have weak spiral patterns. On the other hand, numerical models predict that spiral arms saturate at a certain point, with the amplitude starting to decrease a few Gyr after formation (e.g., \citealp{Fujii2011, Sellwood2022b}). Moreover, the majority of non-dwarf galaxies already exhibit spiral patterns, which may imply that in most of them, the spiral amplitude is decreasing. If this is the case, the possible decrease of $S/T$ over time would be consistent with simulations.
	
	The observed dependence on $\lambda_\text{rf}$ aligns with our findings in \citelinktext{Marchuk2024b}{Paper II}, where we observed that $S/T$ for M~51 decreases from FUV to NIR, and with \citet{Savchenko2020}, where a similar trend was found for a sample of local galaxies, showing that $S/T$ decreases from the $g$ to $i$ band. \citet{Yu2019} measured spiral arm strength at BVRI bands and made a similar conclusion that spiral arms are systematically stronger in bluer bands. However, in our results, the dependence on wavelength is slightly weaker than in both \citelinktext{Marchuk2024b}{Paper II} and \citet{Savchenko2020}. Additionally, the average $S/T$ values reported in \citet{Savchenko2020} are significantly lower than those in this work and in \citelinktext{Chugunov2024}{Paper I}. Note that \citet{Savchenko2020} dealt with the local galaxies, where the weaker spiral pattern can be recognised more easily. Another likely reason for this discrepancy is the difference in methods: in their work, the azimuthally-averaged radial profile of the disc is subtracted before spiral arms are sliced and analysed, whereas, in our decomposition, the disc essentially fits the brightness of the interarm region, resulting in less light being attributed to the disc and more to the spirals. Furthermore, as the placement of slices in their method is based on a visual examination of the image, it is more likely to miss some parts of the spiral arm, leading to an underestimation of its contribution to the luminosity.
	
	\subsubsection{Relative width of the spiral arm}
	\label{sec:width}
	
\begin{figure}[!ht]
		\centering
		\includegraphics[width=0.99\linewidth]{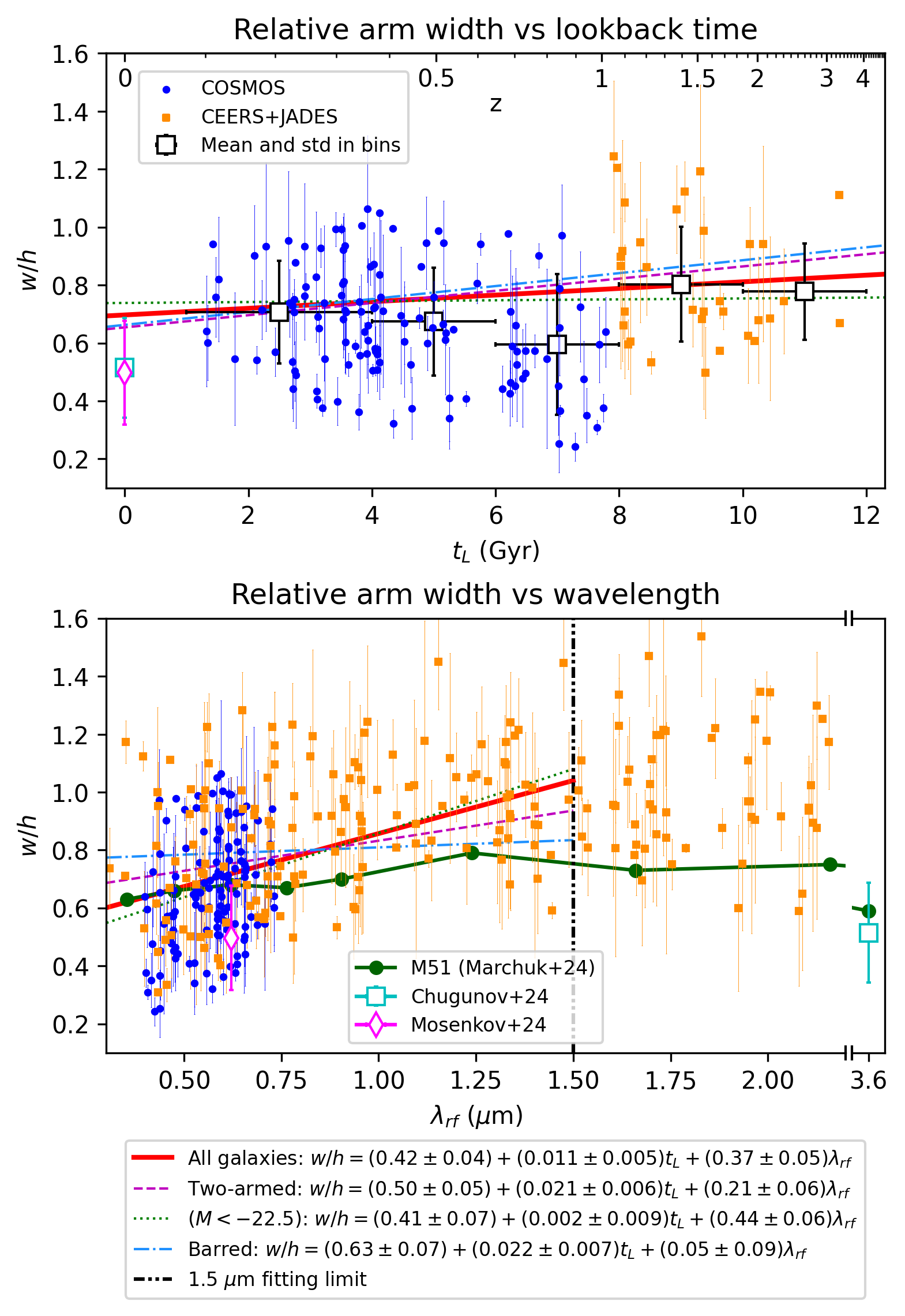}
		\caption{Same as Figure~\ref{fig:ST-time-lambda}, but for the relative width of spiral arm $w/h$ instead of $S/T$. Measurements from~\citet{Chugunov2024, Marchuk2024b, Mosenkov2024} are also shown.}
		\label{fig:wh-time-lambda}
	\end{figure}

	In Figure~\ref{fig:wh-time-lambda}, we present similar diagrams for the relative width of the spiral arm, defined as the width $w$ normalised to the disc's exponential scale $h$. When the dependencies are considered separately, the parameter $w/h$ slightly increases over time, at a rate of 0.11 per 10 Gyr. At the same time, $w/h$ shows a strong dependence on $\lambda_\text{rf}$, increasing by 0.29 per 1~$\mu$m, which is consistent with the trend observed in our \citelinktext{Marchuk2024b}{Paper II}. Overall, the average $w/h$ in the sample is 0.69 with a standard deviation of 0.21. It's important to note that if the spiral arm width is examined using images in a single filter, it appears to decrease with $t_L$, as observed in the COSMOS subsample. However, this dependence reverses when band-shifting effects are accounted for.
	
\begin{figure}[!ht]
		\centering
		\includegraphics[width=0.99\linewidth]{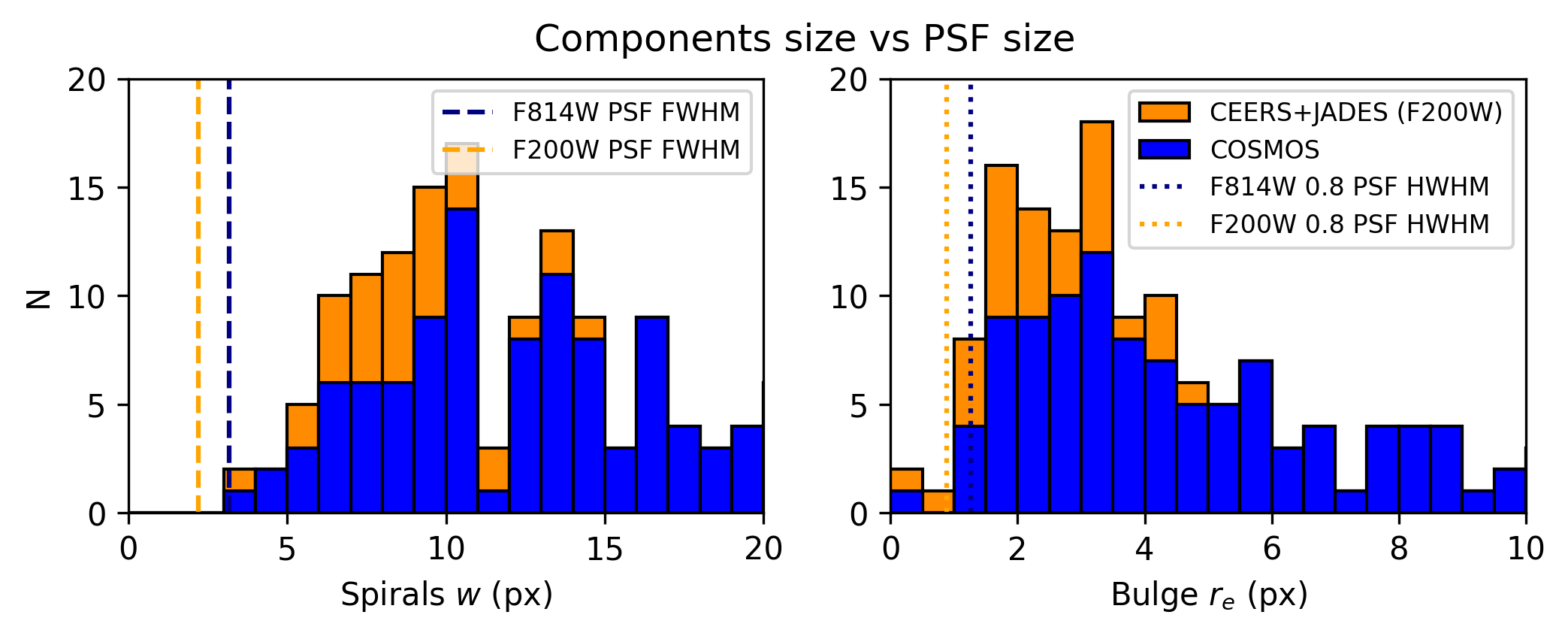}
		\caption{Comparison of spiral arm width $r$ and bulge effective radius $r_e$ in pixels (histograms) and the corresponding PSF size (vertical lines)}
		\label{fig:bulge_and_width}
	\end{figure}

	Since the PSF size varies significantly across different filters (see Table~\ref{tab:bands}), a natural question arises: are we observing a real change in $w$ with $\lambda_\text{rf}$, or is this simply due to the larger PSF sizes in longer-wavelength filters? It is important to note that our decomposition accounts for the PSF, and we measure the intrinsic width of the spiral arm, not the PSF-smeared one. However, in principle, if the PSF FWHM is much larger than the intrinsic width, the latter might not be estimated reliably. Nevertheless, we can confidently rule this out, as the intrinsic width of the spiral arm is, in most cases, significantly larger than the PSF FWHM, as demonstrated in Figure~\ref{fig:bulge_and_width}. Therefore, the PSF does not substantially affect the measured widths.
	
	The result appears to be different for the subsample of barred galaxies, showing a much weaker dependence of $w/h$ on $\lambda_\text{rf}$ (0.05 per $\mu$m) and a stronger dependence on $t_L$. However, this difference should be interpreted with caution, as the barred galaxy subsample is the smallest among those considered. It contains only 52 objects, with just 9 galaxies from the CEERS or JADES surveys. Consequently, multiwavelength data is available for a very limited number of objects, making it challenging to discern relationships between any parameter and $t_L$ or $\lambda_\text{rf}$, as these values are interconnected.
	
	Interestingly, we observe a fairly clear lower boundary for possible $w/h$ values at certain $\lambda_\text{rf}$. For example, at $\lambda_\text{rf}$ between 0.5 and 0.6 $\mu$m, there are no galaxies with $w/h < 0.3$, though several galaxies have $w/h$ just above 0.3. This boundary varies with different $\lambda_\text{rf}$ but remains distinct, with the borderline value increasing as $\lambda_\text{rf}$ increases, similar to the trend in average $w/h$. This observation suggests that some physical mechanism might constrain the width of the spiral arm.
	
		\subsubsection{Relative extent of spiral arms}
	\label{sec:extent}
	
\begin{figure}[!ht]
		\centering
		\includegraphics[width=0.99\linewidth]{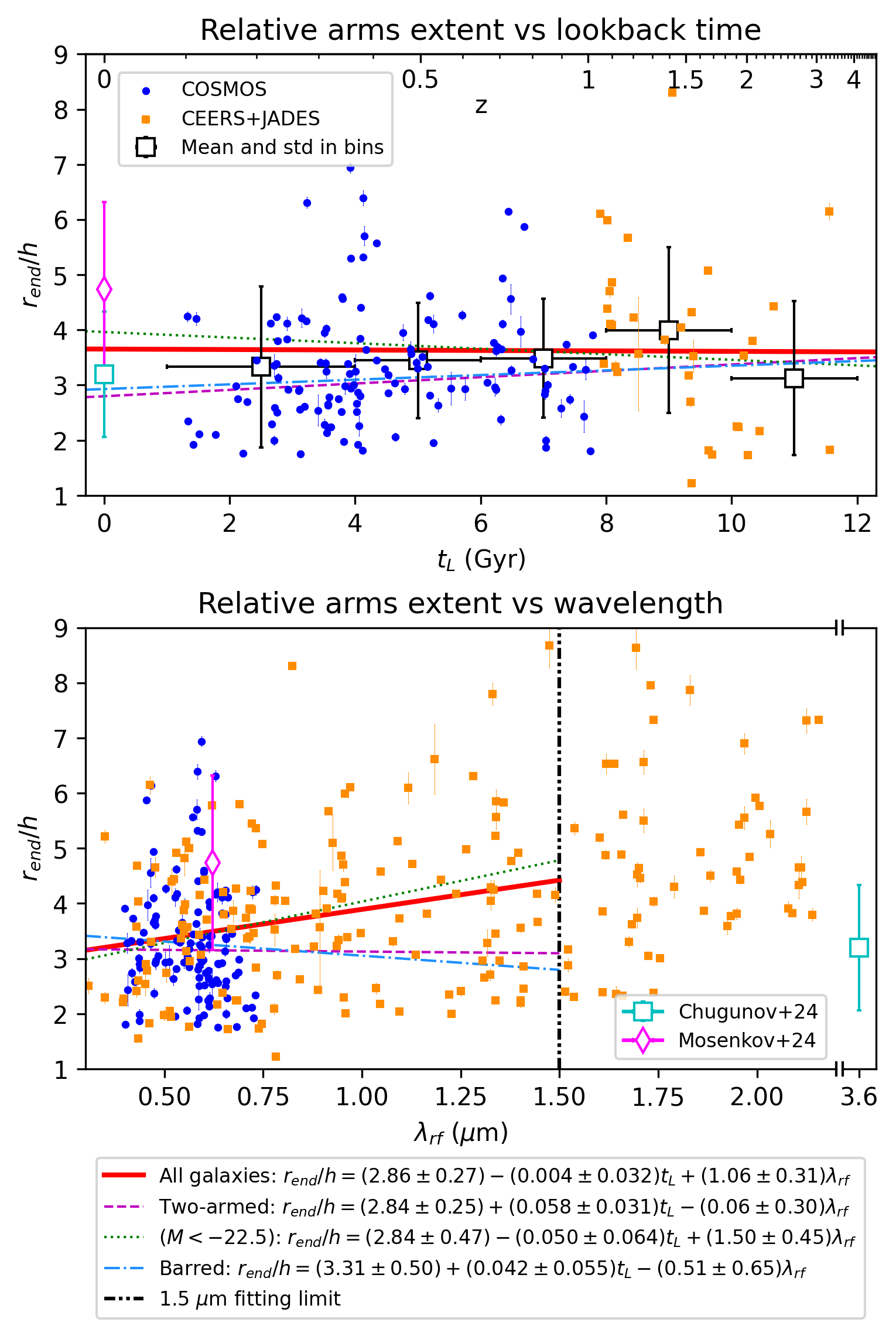}
		\caption{Same as Figure~\ref{fig:ST-time-lambda}, but for the relative extent of spiral arms $r_\text{end}/h$ instead of $S/T$.}
		\label{fig:rendh-time-lambda}
	\end{figure}

	To quantify the extent of spiral arms in a galaxy, we use the highest $r_\text{end}$ parameter among all spiral arms in the galaxy. $r_\text{end}$ itself is one of parameters of the model (see Section~\ref{sec:model}) and is directly obtained from decomposition. This definition is similar to that used in \citet{Mosenkov2024} and indicates how far from the centre the spiral structure extends. We examine the relative arm extent, defined as $r_\text{end} / h$, where $h$ is the disc exponential scale. The relationship of this parameter with $t_L$ and $\lambda_\text{rf}$ is shown in Figure~\ref{fig:rendh-time-lambda}. We found that the extent of spiral arms does not change significantly with time, with an average value of $3.46\times h$. If we consider the optical radius of the galaxy, $R_{25}$, in the given rest-frame instead of $h$, the average value for $r_\text{end} / R_{25}$ is 0.67. Therefore, we trace spiral arms over only a limited part of the disc.
	
	The fitted linear trends depend significantly on the chosen subsample. The dependence of $r_\text{end} / h$ on $\lambda_\text{rf}$ is weaker (if it exists at all) and exhibits much more scatter than the dependence of $w/h$ on $\lambda_\text{rf}$. In principle, one might expect a moderate increase in $r_\text{end} / h$ with $\lambda_\text{rf}$, though weaker than that of $w/h$, because $h$ becomes smaller at longer wavelengths (which also affects $w/h$, see Section~\ref{sec:width}). However, there are no obvious reasons for $r_\text{end}$ to change with wavelength. For instance, \citet{Mosenkov2024} compared the absolute values of spiral arm extents in the $r$ band and near-UV and found no systematic differences.
	
	\subsubsection{Asymmetry}
	\label{sec:asymmetry}
	Now, we will discuss the asymmetry of the spiral structure. First, it is important to clarify that this parameter should not be confused with the skewness parameter of the spiral arm model (see Section~\ref{sec:model}). Asymmetry is a parameter of the spiral pattern as a whole and is not directly related to any parameter of the spiral arm model. To define the spiral structure asymmetry, $A_\text{sp}$, we primarily rely on the widely used definition by \citet{Abraham1996}. However, instead of measuring the asymmetry index for the original image of a galaxy (as is typically done), we apply this method to the model image of the spiral structure only (i.e., the sum of individual spiral arm models)\footnote{Currently, a modified version of this definition is used, which accounts for background noise~\citep{Conselice2000}. However, since we apply this method to a model image that has no noise, there is no need for a more sophisticated approach.}. This also implies that the spiral structure asymmetry index $A_\text{sp}$ is likely to differ significantly from the commonly used asymmetry index $A$ for the same galaxy.
	
\begin{figure}[!ht]
		\centering
		\includegraphics[width=0.99\linewidth]{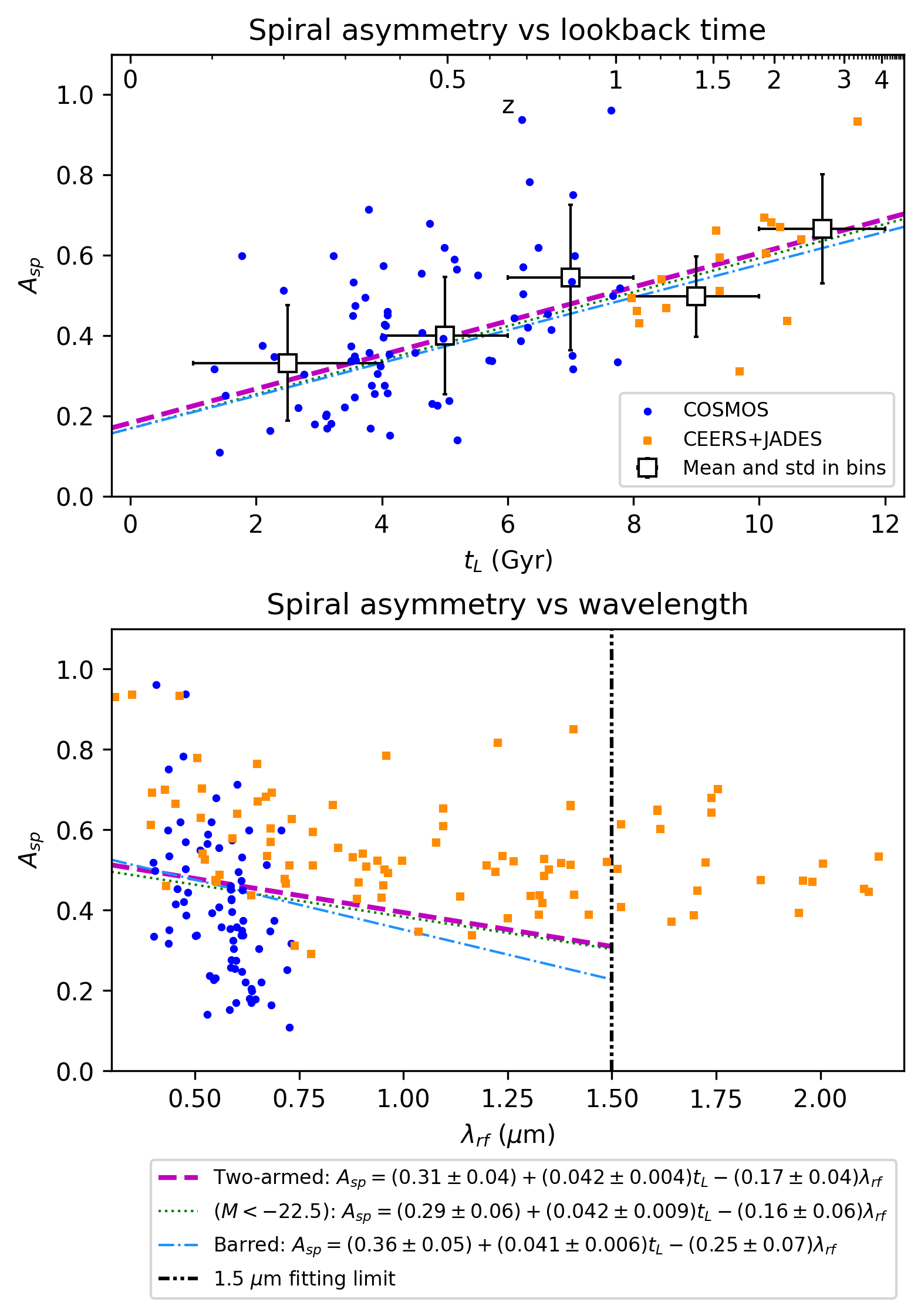}
		\caption{Same as Figure~\ref{fig:ST-time-lambda}, but for the spiral structure's asymmetry $A_\text{sp}$ instead of $S/T$. Only two-armed galaxies are considered.}
		\label{fig:asymm-time-lambda}
	\end{figure}

	The original definition involves the original image of a galaxy, $I_0$ (in our case, the spiral arms model image), and the same image rotated by 180 degrees around the centre, $I_{180}$. The asymmetry index is then defined as $A = \frac{\sum |I_0 - I_{180}|}{2 \sum I_0}$, which essentially measures half the absolute value of the difference between all pixels and their antipodes, relative to the total flux of the galaxy. In Figure~\ref{fig:asymm-time-lambda}, we show how spiral structure asymmetry depends on $t_L$ and $\lambda_\text{rf}$. Since 180-degree rotational symmetry is expected from a two-armed spiral structure, we consider only two-armed galaxies in our analysis, and we expect that a bright and symmetric grand design galaxy will have $A_\text{sp}$ value close to zero.
	
	We observe a strong dependence of $A_\text{sp}$ on both $t_L$ and $\lambda_\text{rf}$. The average $A_\text{sp}$ value for two-armed galaxies is 0.44 (and nearly the same, 0.45 if we consider all galaxies, not just two-armed ones; in both cases, the standard deviation is 0.18). Spiral structures become significantly less symmetric at higher redshifts, with $A_\text{sp}$ increasing with $t_L$ at a rate of 0.04 per Gyr. At the same time, spiral structures appear more symmetric at longer wavelengths: $A_\text{sp}$ decreases with $\lambda_\text{rf}$ by $-0.17$ per $\mu$m. We examined the correlation between the signal-to-noise ratio of the original image and $A_\text{sp}$ to ensure that the observed trend is not influenced by variations in the image's ability to constrain the model.
	
	The spiral structure asymmetry is distinct from the overall galaxy asymmetry, but the two are likely connected, as spiral arms are the primary large-scale non-axisymmetric component in a galaxy. Since the asymmetry index is higher in galaxies that have undergone mergers, and merger rates were higher in the past \citep{Conselice2007, Conselice2014}, it is reasonable to expect that spiral galaxies at higher redshifts are more asymmetric. \citet{Yao2023} found that the commonly used asymmetry index is smaller at longer wavelengths, although they did not observe a strong dependence on redshift. Meanwhile, more distant galaxies tend to be clumpier up to $z$ = 1--3 \citep{Shibuya2016}, and since clumps, which are locations of star formation, can be associated with spiral arms, higher clumpiness may indicate greater irregularity and asymmetry in spiral arms. The decrease of $A_\text{sp}$ with wavelength may be related to the increase in spiral arm width with wavelength, as spiral arms rotated by 180 degrees are more likely to overlap if they are wider.
	
	Since we use model images that are corrected for PSF, the related observational effects, such as decreasing spatial resolution at longer wavelengths and higher redshifts, are excluded from the observed trends.
	
	\section{Validation}
	\label{sec:validation}
	
	Several factors could potentially skew the results presented in Section~\ref{sec:results}. These include the incompleteness of the sample, which leads to a lack of faint galaxies at large $z$ (see Figure~\ref{fig:basic_data}), and possible challenges in measuring the full extent of spiral arm parameters. Another problem to consider is model complexity, which can potentially lead to instability in parameter measurements and introduce systematic bias into the results. In this section, we examine how these effects might impact our findings.
	
	\subsection{Possible dependence on luminosity}
	\label{sec:completeness}
	The average luminosity of galaxies in our sample varies significantly with redshift (see Figure~\ref{fig:basic_data}), indicating that the sample is incomplete in terms of absolute magnitude. This variation could lead to a potential issue: if a parameter depends on galaxy luminosity, the observed dependence of that parameter on $t_L$ may actually reflect luminosity dependence rather than evolutionary changes over time. In this section, we address this concern.
	
	A straightforward way to account for the sample's incompleteness is to apply a limiting absolute magnitude and ensure that galaxies as faint as this magnitude are represented at all redshifts. As described in Section~\ref{sec:data}, we use $M_\text{F814W} = -22.5$ as the limiting magnitude, resulting in 63 galaxies that are brighter than this limit. While this limit is somewhat arbitrary, even if this subsample remains incomplete at this magnitude, its degree of incompleteness is less than that of the full sample. Therefore, if parameters do vary with luminosity, any differences in results between the bright galaxy subsample and the full sample should be noticeable. However, as seen in Section~\ref{sec:results}, the trends remain largely the same when considering only bright galaxies.
	
	A more detailed approach to examining the effects of the sample's incompleteness is to measure the dependencies of parameters on absolute magnitude, $M_\text{F814W}$. We can analyze the dependence on $M_\text{F814W}$ while simultaneously considering $M_\text{F814W}$, $t_L$, and $\lambda_\text{rf}$, similar to the method we used to distinguish between evolutionary and band-shifting effects in Section~\ref{sec:discerning}. We conduct this analysis and also examine correlations with $M_\text{F814W}$ alone. We find that only one parameter consistently correlates with $M_\text{F814W}$ alone and shows changes with $M_\text{F814W}$ when $t_L$ and $\lambda_\text{rf}$ are also considered: $w/h$. On average, $w/h$ decreases by 0.06 per magnitude toward brighter galaxies, and the dependencies on $t_L$ and $\lambda_\text{rf}$ remain roughly the same for $w/h$. Interestingly, a similar but even stronger trend is observed in our \citelinktext{Chugunov2024}{Paper I} data, indicating that brighter galaxies tend to have relatively narrower spiral arms in the 3.6~$\mu$m band. Conversely, we do not observe any consistent variation of $S/T$ with $M_\text{F814W}$, whereas in \citelinktext{Chugunov2024}{Paper I}, we clearly found that brighter galaxies have higher $S/T$. This difference is likely attributable to the fact that \citelinktext{Chugunov2024}{Paper I} dealt with a sample of fainter galaxies than those in this study, suggesting that the $S/T$ dependence on luminosity may exist only within a limited range of absolute magnitudes or in certain filters.
	
\begin{figure*}[!ht]
		\centering
		\includegraphics[width=0.99\linewidth]{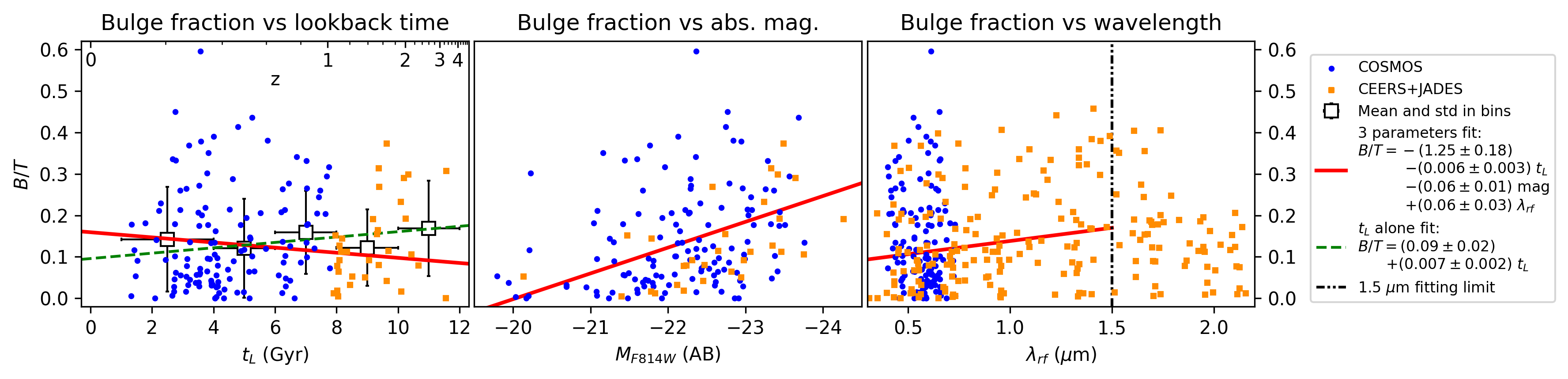}
		\caption{Bulge-to-total ratio vs. $t_L$ (left), $M_\text{F814W}$ (centre), and $\lambda_\text{rf}$ (right). The green dashed line represents the linear fit to the dependence of $B/T$ on $t_L$ alone, while the red line represents the projection of the trilinear fit (see legend) to the dependence of $B/T$ on $t_L$, $M_\text{F814W}$, and $\lambda_\text{rf}$.}
		\label{fig:bulge_relations}
	\end{figure*}

	We can also apply this method to any other measured parameter of the model to assess its plausibility and efficiency. Here, we consider the bulge fraction in the total luminosity of a galaxy, $B/T$, as measured from our decomposition. Although $B/T$ is not the main focus of our work\footnote{For detailed information on $B/T$ measurement in decompositions that include spiral arms, refer to our \citelinktext{Chugunov2024}{Paper I} and \citelinktext{Marchuk2024b}{II}.}, we highlight that $B/T$ is a relatively well-studied parameter, with its variations with redshift, wavelength, and galaxy luminosity already known. In Figure~\ref{fig:bulge_relations}, we present three diagrams of $B/T$ against $t_L$, $M_\text{F814W}$, and $\lambda_\text{rf}$.
	
	If we analyse the dependence of $B/T$ simply on $t_L$, we find that more distant galaxies have higher $B/T$, which, if considered alone, would suggest that $B/T$ decreases over time. Such a conclusion would contradict most modern observations and simulations \citep{Bruce2014, Brooks2016, Sachdeva2017}, as well as theoretical predictions that bulges grow through mergers of galaxies and dynamical evolution \citep{Hopkins2010}. However, when we fit $B/T$ as a trilinear function of $t_L$, $\lambda_\text{rf}$, and $M_\text{F814W}$, a different picture emerges: $B/T$ decreases at larger $t_L$, as expected (if luminosity and wavelength remain constant), and this is accompanied by a strong increase of $B/T$ in more luminous galaxies and a moderate increase at longer wavelengths. In other words, the apparent increase of $B/T$ with $t_L$ inferred from the simple linear fit is actually due to the fact that more distant galaxies are, on average, brighter, and $B/T$ is significantly higher for bright galaxies. Additionally, $\lambda_\text{rf}$ is higher for CEERS and JADES samples than for COSMOS in most filters. This behavior observed in the trilinear fit aligns well with theoretical predictions and observations. It is well known that more massive (and therefore brighter) galaxies tend to be more bulge-dominated \citep{Khochfar2011}, and observed $B/T$ should also be larger at longer wavelengths, as older stellar populations are concentrated in the bulge (see, for example, \citealp{Gong2023} and our \citelinktext{Marchuk2024b}{Paper II}). For the vast majority of objects in our sample, the effective radius $r_e$ of their bulge is greater than $0.8\times$HWHM of the PSF (see Figure~\ref{fig:bulge_and_width}), which meets the requirement for reliably extracting bulge parameters \citep{Gadotti2009}. However, we lack information on the presence and luminosity of AGN in the galaxies in our sample, which could potentially introduce biases in the measured bulge parameters.
	
	Returning to the parameters of spiral arms, we can conclude that their properties do not strongly depend on luminosity within our sample. Therefore, the sample's incompleteness does not significantly impact our results.
	
	\subsection{Change of resolution and surface brightness over redshift}
	\label{sec:change_of_resolution}
	As mentioned in Section~\ref{sec:length}, the spatial resolution of a galaxy decreases as its distance increases (although at $z \approx 1.7$, this trend reverses due to cosmological effects, see \citealp{Melia2018}). Another effect is cosmological surface brightness dimming, which results in the apparent surface brightness of a galaxy decreasing proportionally to $(1+z)^{-3}$ in AB magnitudes (not to the power of $-4$ because AB magnitudes correspond to erg s$^{-1}$ cm$^{-2}$ Hz$^{-1}$; see \citealp{Whitney2020} for details and derivation).
	
	Therefore, it is natural to expect that the results of decomposition could vary for similar galaxies at different redshifts. In other words, some changes in the measured parameters with $z$ could potentially be attributed to these factors. We need to determine whether our findings are influenced by this possible effect.
	
	Decreasing spatial resolution primarily impacts the inner components of a galaxy, as these regions are already small in size. Specifically, the inner parts of spiral arms become more difficult to discern. Another effect, cosmological dimming, is likely to cause us to miss the faintest peripheral parts of distant galaxies. This could result in the outer parts of spiral arms also going unnoticed, leading to an underestimation of azimuthal lengths of the spiral arms (Section~\ref{sec:length}) and their extent (Section~\ref{sec:extent}). Indeed, \citet{Mosenkov2024} measured these parameters in deep DESI images and found that, compared to less deep SDSS images, the observed arm length and extent increase significantly (depending on galaxy type; for multi-armed galaxies, the relative increase is 18\% for both parameters). We are going to test if these effects are significant by preparing a set of artificially redshifted images and analysing them in the same way as the original ones in the following subsection.
	
	\subsubsection{Artificial redshifting}
	
	In practice, we can test our conjectures by creating a sample of artificially redshifted images of galaxies and analysing how the parameters change with $z$ for the same galaxy~\citep{Kuhn2023, Martinez-Garcia2023}. The decreasing spatial resolution can be effectively modelled by resampling the image and convolving it with an adjusted PSF, while the decreasing surface brightness can be simulated by reducing the overall brightness in the image and applying the original background noise, or by adding appropriate noise to the model image. Another aspect of artificial redshifting, specifically the change in $\lambda_\text{rf}$, is challenging to reproduce in full because the contributions of different galaxy components vary with wavelength~\citep{Yu2023}. For instance, spiral arms appear more clumpy at higher wavelengths, but it's worth noting that spiral arm length and extent --- the parameters we are primarily concerned with --- are not strongly affected by $\lambda_\text{rf}$ in any case.
	
\begin{figure}[!ht]
		\centering
		\includegraphics[width=0.99\linewidth]{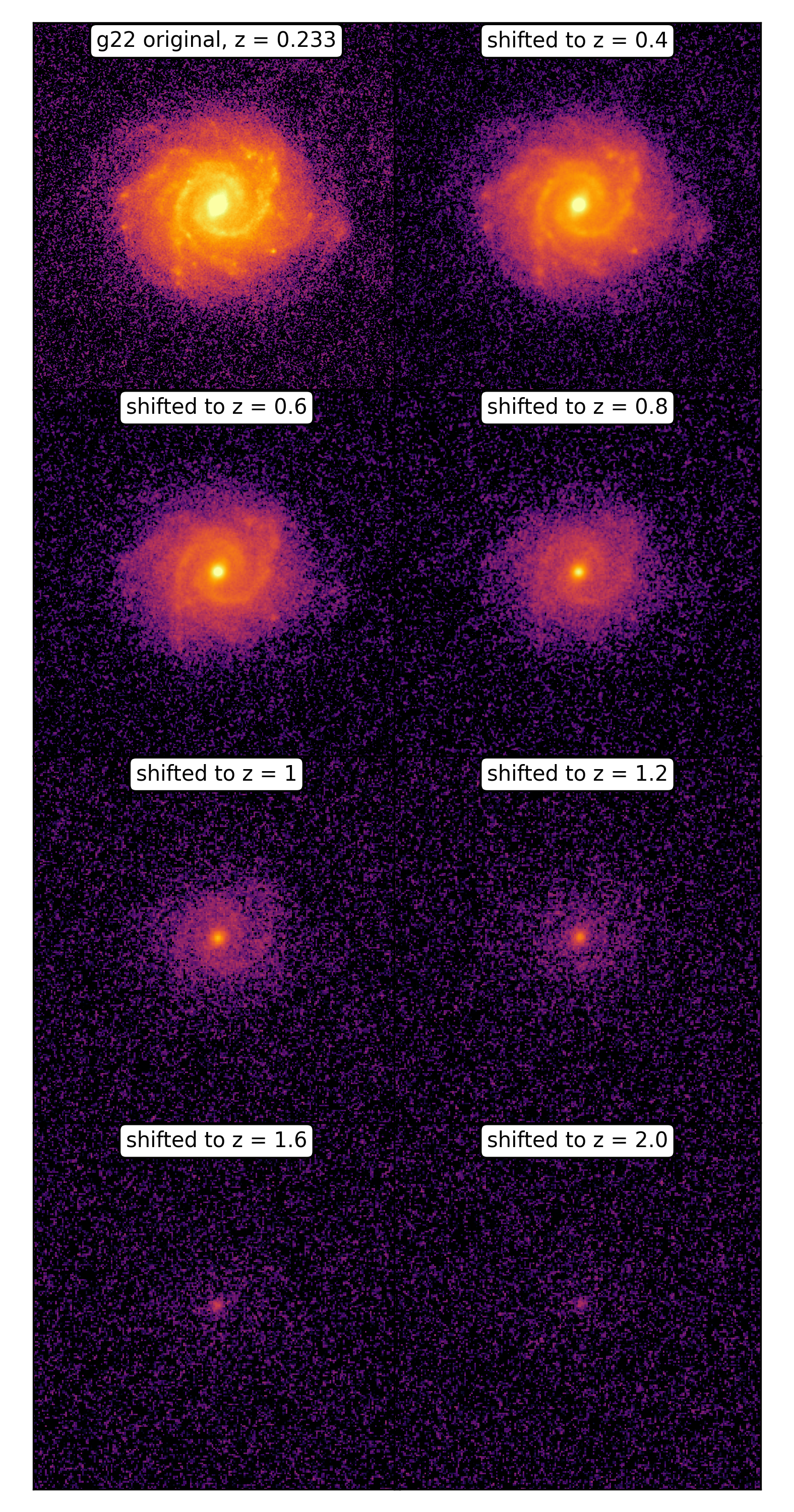}
		\caption{An example of a galaxy at $z = 0.233$ (original image shown in the upper-left) being artificially redshifted to different redshifts. For additional reference, see Figure~4 in \citet{Kuhn2023}, where artificial redshifting of spiral galaxies is demonstrated.}
		\label{fig:redshifting}
	\end{figure}

	To perform artificial redshifting, we select the closest but brightest galaxies from the COSMOS subsample and adjust the spatial resolution, PSF, and relative noise to match the parameters of COSMOS images at several different redshifts. We also apply K-correction based on the photometric measurements from \citet{Weaver2022}, in the same manner that we obtained F814W magnitude (Section~\ref{sec:data}). An example set of redshifted images is shown in Figure~\ref{fig:redshifting}. We analyse redshifted images up to $z = 1$ because, as we will demonstrate, the quality of COSMOS images at $z = 1$ is comparable to the quality of CEERS and JADES images at $z \approx 2.3$.
	
	We remind that our sample consists of images of varying quality: the HST COSMOS survey has significantly lower photometric depth compared to both JWST surveys. We measured the limiting surface brightness depth for COSMOS as 29.8 AB(F814W) mag/arcsec$^{-2}$, whereas for CEERS, it ranges from 31 to 32 AB mag/arcsec$^{-2}$ in different filters (see Table~\ref{tab:bands}). JADES images are slightly deeper. For the COSMOS subsample, the highest redshift used is approximately $z = 1$. To estimate the redshift at which the effective depth of CEERS images matches that of COSMOS images at $z = 1$, we consider that cosmological dimming is proportional to $(1 + z)^{-3}$ in AB magnitudes. Thus, we can estimate that the dimming in AB magnitudes at $z = 2.3$ (the highest redshift in our sample, except for two galaxies) is stronger than at $z = 1$ by a factor of $\left(\frac{2.3 + 1}{1 + 1}\right)^3 \approx 4.4$, which is nearly 1.6 mag, comparable to the difference between the COSMOS and CEERS depths. 
	
	In addition to that, for most CEERS galaxies and filters, $\lambda_\text{rf}$ is higher than for COSMOS galaxies, but for the F814W filter at $z = 1$, it is roughly the same as for the F150W filter at $z = 2.3$, approximately $0.4 \mu$m. Finally, the spatial resolution changes only slightly between $z = 1$ and $z = 2.3$. As a result, we conclude that the quality of COSMOS images at $z = 1$ approximately corresponds to the quality of CEERS images at $z = 2.3$, and there is no need to analyse redshifts of COSMOS galaxies beyond $z = 1$. This corresponds well with the observation that $z \approx 1$ is the highest redshift at which spiral galaxies from the COSMOS sample are observed, while $z \approx 2.3$ is the highest for CEERS and JADES samples, except for two galaxies. At higher redshifts, the image quality in these surveys generally becomes too poor to include spiral galaxies, albeit they exist at these redshifts \citep{Kuhn2023}, in our samples.
	
\begin{figure}[!ht]
		\centering
		\includegraphics[width=0.99\linewidth]{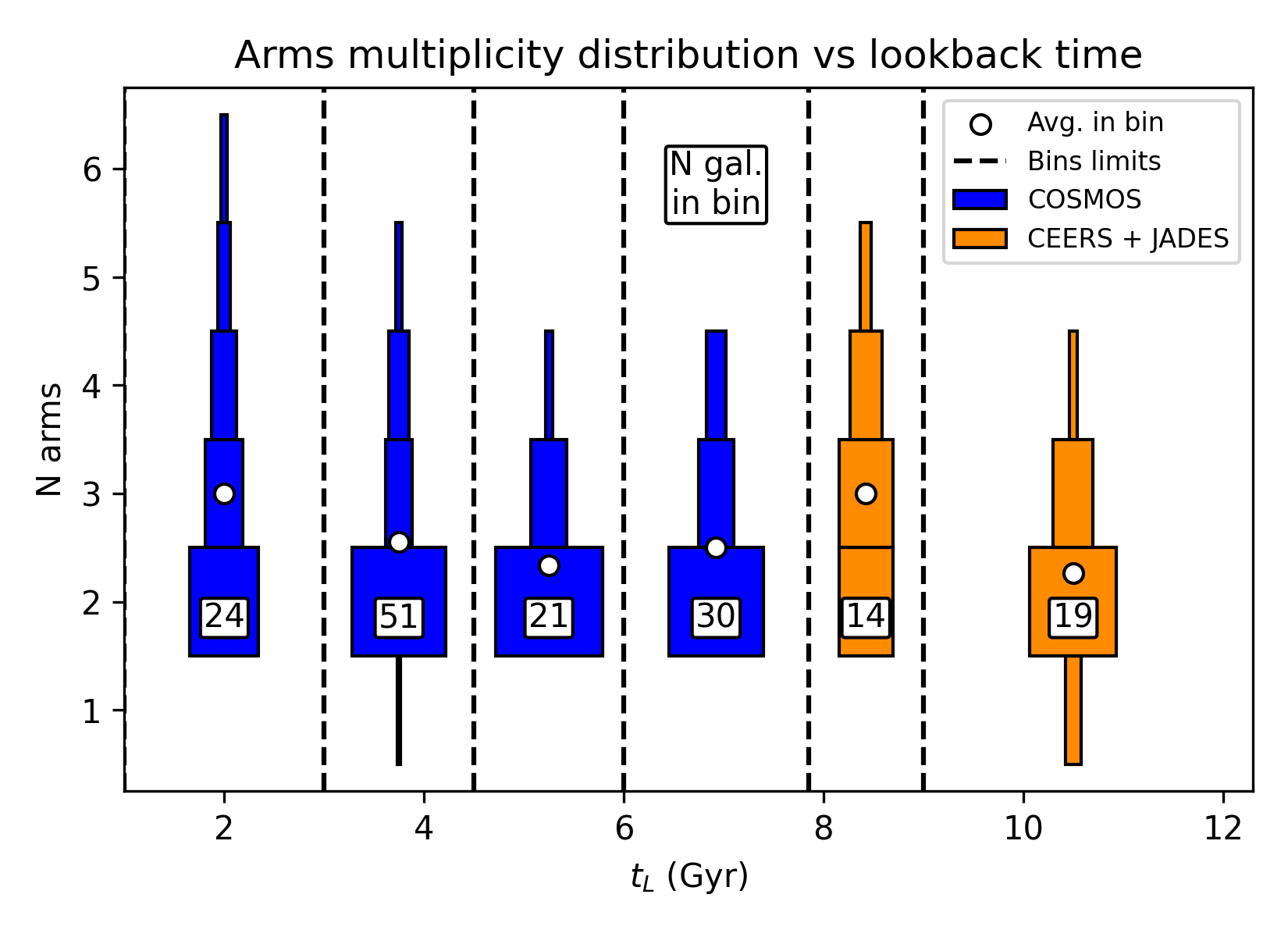}
		\caption{In this illustration, each vertical histogram shows the distribution of galaxies by the number of arms within the corresponding $t_L$ bin, with bin borders marked by dashed lines. The row width in each histogram represents the relative frequency of galaxies with $N$ arms in each bin. The average number of spiral arms in each bin is indicated by a white circle, and the number in the box shows the total number of galaxies in the corresponding bin. Note the increase in the average number of spiral arms after the last COSMOS bin, which is due to the improved image quality in the CEERS and JADES data.}
		\label{fig:N_arms}
	\end{figure}

	The difference in photometric depths between the samples also suggests that at $z \approx 1$, where the CEERS and JADES samples replace COSMOS, there is an abrupt increase in image quality for the actual images in our samples. This is also noticeable when examining the average number of spiral arms versus $z$ (see Figure~\ref{fig:N_arms}), which decreases up to $z = 1$, then abruptly increases, and starts decreasing again at $z > 1$. This pattern is likely due to the fact that when the signal-to-noise ratio is poor, only the most prominent spiral arms are detected. For example, in Figure~\ref{fig:redshifting}, the same galaxy at $z = 0.233$ is artificially redshifted; at $z \approx 1$, only the two main spiral arms are visible, whereas multiple spiral arms are noticeable in the original image.
	
	Despite this, we do not observe a similar discontinuous behaviour near $z \approx 1$ for any of the previously mentioned parameters, which we might expect if the parameters were actually dependent on image quality—most notably for $l_\psi$ and $r_\text{end} / h$. The only apparent exception is $w/h$ (Figure~\ref{fig:wh-time-lambda}); however, it is important to note that the data points for the CEERS and JADES subsamples on the left part of Figure~\ref{fig:wh-time-lambda} correspond to much larger $\lambda_\text{rf}$ values than those for the COSMOS subsample. At the same time, $w/h$ is a parameter with a very strong dependence on $\lambda_\text{rf}$, which explains this apparent discontinuity. This indicates that the sudden change in image quality near $z \approx 1$ does not significantly affect the measured parameters. Additionally, we note that the parameters of galaxies from COSMOS and from CEERS and JADES are mostly similar over the overlapping part of the $\lambda_\text{rf}$ range; see, for example, Figs.~\ref{fig:wh-time-lambda} and~\ref{fig:rendh-time-lambda}.
	
	Therefore, this provides some evidence that variations in image quality with $z$ do not significantly affect our findings. However, in the next section, we will conduct a more thorough analysis to confirm this.
	
	\subsubsection{Analysis of artificial redshifted images}
	
	Subsequently, we performed decomposition for 10 galaxies artificially redshifted from $z\approx 0.25$ to several $z$ values up to 1, using the same method as for the original images, and measured the same set of parameters. We visually identified the number and approximate location of spiral arms in each case (see Section~\ref{sec:decomposition}) and fitted their parameters. As expected, the number of observed spiral arms decreases with increasing $z$, primarily due to the apparent disappearance of the weaker spiral arms.
	
\begin{figure}[!ht]
		\centering
		\includegraphics[width=0.99\linewidth]{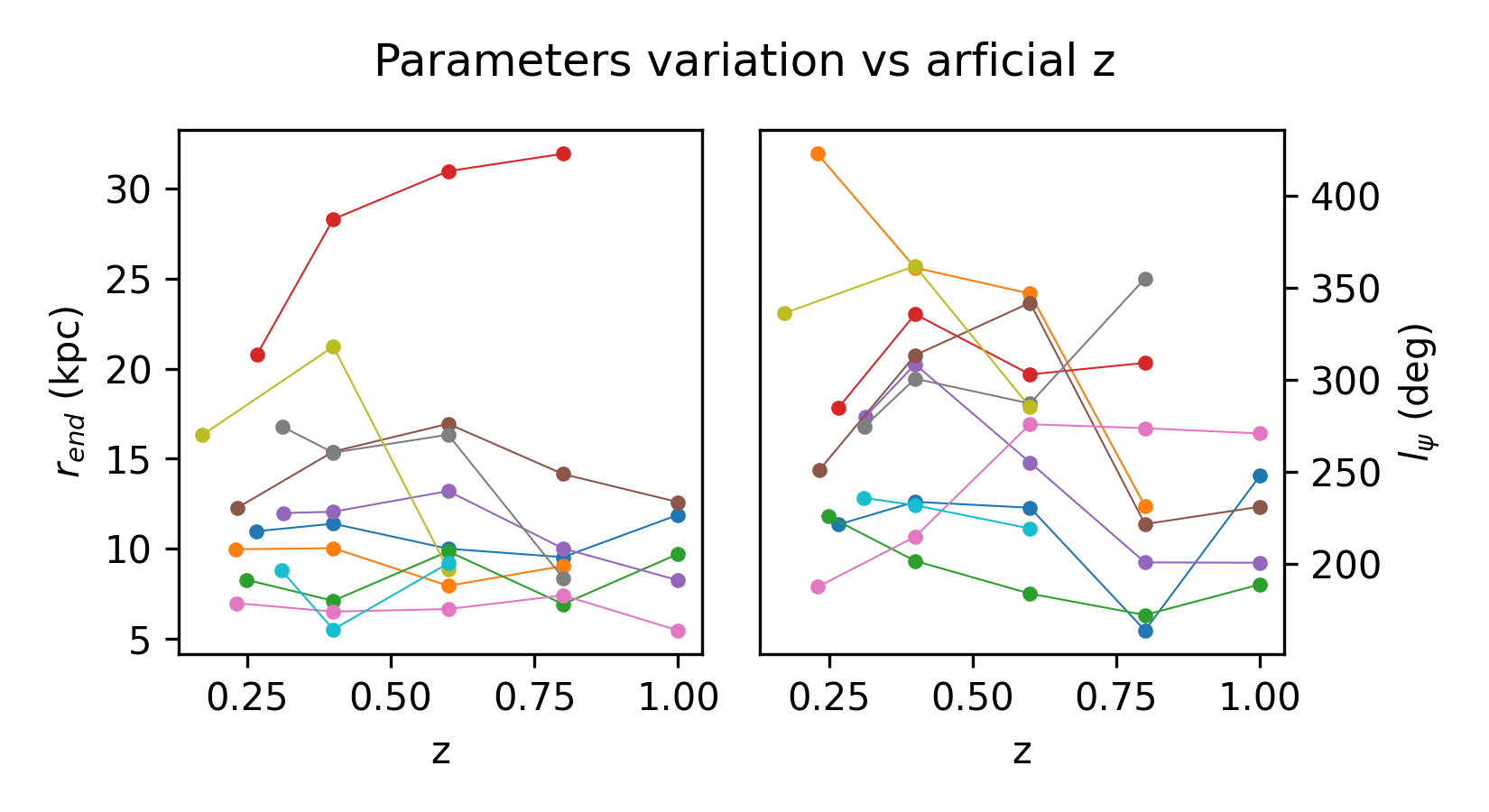}
		\caption{Behaviour of the relative extent of spiral arms, $r_\text{end}$ (left), and azimuthal length, $l_\psi$ (right), for original and artificially redshifted images as a function of $z$. Each coloured line represents a single galaxy from COSMOS at the original and artificial $z$ values where decomposition was performed, showing the measured values of $r_\text{end}$ and $l_\psi$.}
		\label{fig:validation_decomp}
	\end{figure}

	It appears that a variety of factors can influence the measured parameters, making the behaviour of spiral arm length $l_\psi$ and extent $r_\text{end}$ far from straightforward. In Fig~\ref{fig:validation_decomp}, we illustrate how these measured parameters vary with $z$ for both the original and artificially redshifted images of individual galaxies.
	
	There are several reasons why the length and extent of spiral arms vary with redshift. In some cases, the behaviour aligns with what was initially discussed and expected in Section~\ref{sec:change_of_resolution}: spiral arms become shorter with increasing redshift as their inner parts become resolved too poor and peripheries become harder to discern. However, even when a spiral arm appears visually shorter at the periphery, its outer parts can still be recovered through decomposition.
	
	In some instances, the observed length can increase with increasing $z$, as seen in the case of galaxy g22, which is also shown in Figure~\ref{fig:redshifting}. This galaxy, located at $z = 0.233$, has multiple spiral arms, and we modelled 6 of them in the original image. As the galaxy is redshifted to $z = 0.6$, only 4 arms remain recognisable, and at $z = 1$, only the 2 main spiral arms are still visible. Naturally, the smallest and faintest spiral arms disappear first, while the main ones persist up to $z = 1$. Even though their apparent length changes with $z$, the disappearance of weaker and shorter arms leads to an increase in the average length. In more extreme cases, spiral arms that appear separate at smaller $z$ can become too smooth at larger $z$ and merge into a single continuous arm, which may naturally increase the average length.
	
	Finally, with decreasing resolution and increasing noise, the bright parts and segments of the galactic structure may appear rearranged in different configurations, resulting in a completely altered spiral structure shape.
	
	All these effects result in $r_\text{end}$ and $l_\psi$ either decreasing or increasing with redshift, varying uniquely for each galaxy. While the former parameter seems to be less affected by the image quality, the decrease in $l_\psi$ is typically more pronounced. Therefore, we must attempt to measure this effect in a quantitative way.
	
\begin{figure}[!ht]
		\centering
		\includegraphics[width=0.99\linewidth]{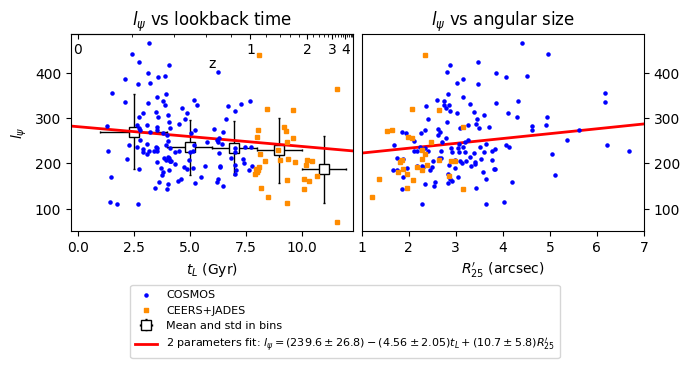}
		\caption{Azimuthal length $l_\psi$ vs. $t_L$ (left) and $R_{25}'$ (right). The red line represents the projection of the bilinear fit (see legend) to the dependence of $l_\psi$ on both $t_L$ and $R_{25}'$.}
		\label{fig:length-time-r25}
	\end{figure}

	We observe that the strongest factor contributing to the decrease in $l_\psi$ with increasing artificial $z$ is the inability to discern the central parts of the spiral structure when spatial resolution is poor. A significant portion of the azimuthal length of an individual spiral arm often resides in the relatively small region near the centre of a galaxy, which can be completely lost as redshift increases. To quantify the degree of resolution, we use the angular optical radius $R_{25}'$ of a galaxy, which directly correlates with the number of pixels occupied by the galaxy on the image, since the pixel size is consistent across all images we analyse, specifically 0.03 arcsec/pix. We perform a bilinear fit of $l_\psi$ as a function of $t_L$ and $R_{25}'$. In Figure~\ref{fig:length-time-r25}, one can see that a dependence on $R_{25}'$ exists, indicating that angularly larger galaxies tend to have longer spiral arms. However, the dependence on $t_L$ remains, albeit weaker than it initially appeared in Section~\ref{sec:length}: $l_\psi$ decreases by 5 deg per Gyr.
	
\begin{figure}[!ht]
		\centering
		\includegraphics[width=0.99\linewidth]{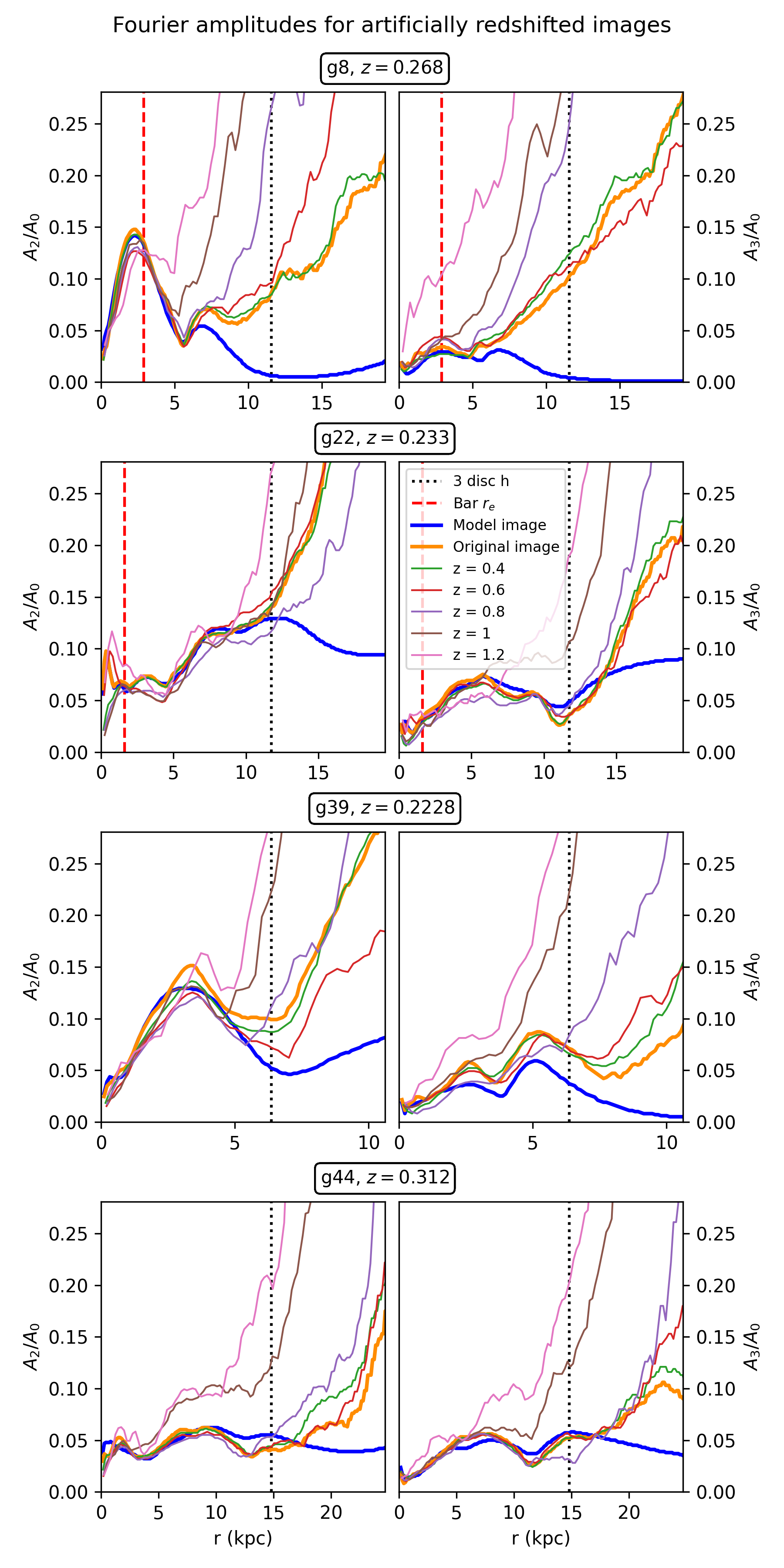}
		\caption{Fourier mode amplitudes $A_2$ and $A_3$ relative to $A_0$ of galaxies as a function of radius, shown as a function of redshift. Four galaxies are presented: g8, g22 (also shown in Figure~\ref{fig:redshifting}), g39, and g44, from top to bottom. The diagrams for relative amplitudes $A_2$ and $A_3$ are in the left and right columns, respectively. Amplitudes are shown for the original image, artificially redshifted images up to $z = 1.2$, and the model image. The bar effective radius is indicated with a dashed line, where present, and three times the exponential disc scale is marked with a dotted line, representing a typical radius of spiral arm truncation.}
		\label{fig:validation_fourier}
	\end{figure}

	To further validate our results, we performed a 2D Fourier analysis using \verb|P2DFFT| \citep{Hewitt2020}, with the method itself described in \citet{Davis2012}. In Figure~\ref{fig:validation_fourier}, we show an example of Fourier mode amplitudes as a function of radius for the original images, artificially redshifted images, and model images. We consider the amplitudes of modes $A_2, A_3, A_4$ relative to the axisymmetric component amplitude $A_0$. Throughout most of the disc, the amplitudes of the modes in the original and model images are similar, as expected, and they begin to diverge at the periphery where the amplitude of the modes in the original image starts to increase rapidly. This divergence is caused by the decreasing signal-to-noise ratio, with the divergence point corresponding to the radius beyond which the noise contribution to the amplitude overtakes that of the spiral arms.
	
	Artificially redshifted images have a lower signal-to-noise ratio, so they are expected to diverge from the model image amplitudes at smaller radii. This is indeed the case, but up to $z \approx 1$, the shift in the point of divergence is minimal. The same pattern is observed for other galaxies, suggesting that the ends of the main spiral arms in COSMOS images begin to disappear near $z \approx 1$.
	
	Therefore, most parameters of spiral arms measured in our study, do not depend significantly on the observational effects of decreasing image depth and varying spatial resolution, and only azimuthal length depends moderately on them.
	
	\subsection{Parameter stability}
	Another problem to consider is the complexity of our model. Given the remoteness of the galaxies in the sample, it is essential to evaluate whether the high number of parameters in our model is justified. We need to ensure that this complexity does not result in unstable parameter behaviour or introduce systematic biases into our results.
	
	Determining the optimal number of parameters for a spiral arm model is a non-trivial task, and we plan to explore this further in a separate study (Chugunov \& Marchuk, in prep.). However, for the current sample, we can assess the stability of the parameters by examining how their measurements change when the model is simplified. Specifically, we evaluate how the pitch angle estimate is affected when the spiral arm shape function (Equation~\ref{eq:r_psi}) is simplified, and how the measured width changes when the function defining the perpendicular profile (Equation~\ref{eq:I_bot}) is simplified. To do this, we randomly select 10 galaxies from our COSMOS subsample and perform decomposition with some parameters fixed at some ``default'' values, thereby simplifying the model. We then compare the parameter values obtained using this ``simplified'' arm model with those derived from the original, more complex model.
	
	First, we test the simplified shape function of a spiral arm (Equation~\ref{eq:r_psi}) with $n$ ranging from 1 to 3. In most cases, the resulting spiral arms are nearly indistinguishable from those described by the original model. However, there are exceptions, most notably in the case of galaxy g26 (see Figure~\ref{fig:COSMOS_mosaic_1}). This galaxy features long spiral arms with strongly varying pitch angles, and the simplified shape function accurately represents the shape of one arm only within a limited range of azimuthal angles. We then measure the pitch angles of the simplified spiral arms using the same method applied to our main results (Section~\ref{sec:pitch}). Comparing these measurements, we find only a small difference, with an average absolute value of 1.9 degrees and no significant bias --- this difference is smaller than that arising when different methods are compared. Moreover, in some cases, such as the aforementioned g26, the difference is justified because the simpler model fails to capture part of the spiral arm.
	
\begin{figure}[!ht]
		\centering
		\includegraphics[width=0.99\linewidth]{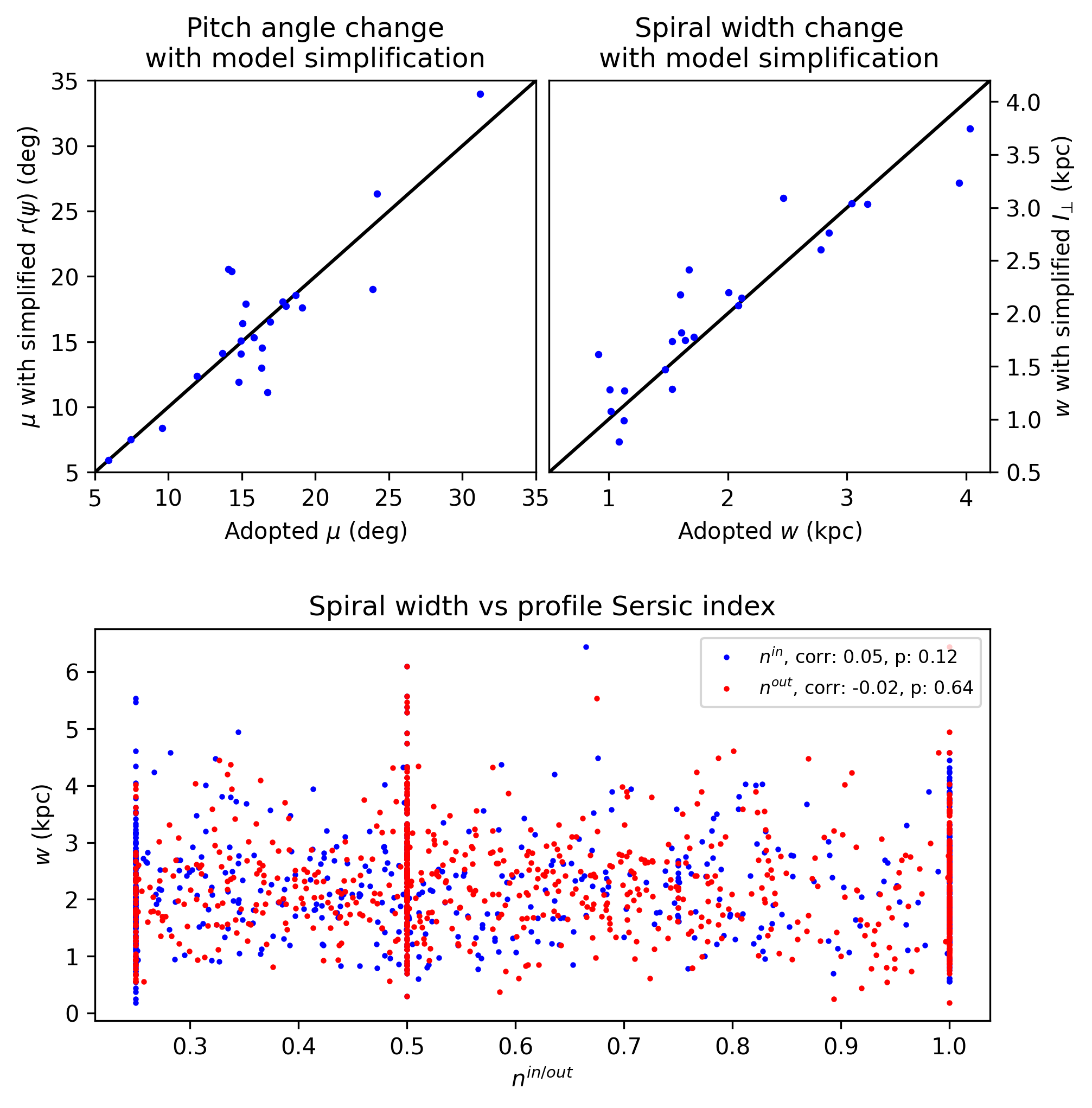}
		\caption{The top-left panel compares pitch angles measured using the original and simplified models of spiral arms. The top-right panel compares widths measured with the original and simplified models of spiral arms. The bottom panel shows the measured widths displayed against the measured $n^\text{in/out}$ for each spiral arm. The correlation coefficients and p-values for $n^\text{in}$ (blue) and $n^\text{out}$ (red) are shown in the legend. The points are clustered at certain $n^\text{in/out}$ values because 0.25 and 1 were used as lower and upper limits for fitting. In some cases, $n^\text{in/out}$ was fixed at 0.5. This was done in less than 20\% of cases, typically when non-fixed $n^\text{in/out}$ produced spiral arms with implausible profiles.}
		\label{fig:params_stability_test}
	\end{figure}

	Next, we test the simplified profile across the spiral arm (Equation~\ref{eq:I_bot}). By fixing both $n^\text{in/out}$ to 0.5, we effectively reduce the asymmetric S\'ersic profile to an asymmetric Gaussian. Since different spiral arms have different $n^\text{in/out}$, their measured widths are not necessarily tied to the same part of their profile. In principle, if $n^\text{in/out}$ is found to correlate with width, it could introduce some systematic bias. To evaluate this, we compare not only the differences between the simplified and original models but also examine the correlation between the measured $n^\text{in/out}$ values and $w$. Overall, we observe some scatter in the measured widths around the 1:1 relation; however, no significant correlation is found between $w$ and $n^\text{in/out}$. This lack of correlation is likely because we use the FWHM as the measure of spiral arm width rather than the effective radius. Therefore, even if systematic variations in $n^\text{in/out}$ occur, whether due to physical or observational factors, they likely do not introduce any bias into our width measurements. The results of our comparisons are shown in Figure~\ref{fig:params_stability_test}.
	
	\section{Discussion}
	\label{sec:discussion}
	There are very few studies that have examined the spiral structure of distant galaxies, leaving limited data for comparison with our results. Additionally, theoretical works on this subject are scarce and primarily focus on a single parameter of spiral structure, namely the pitch angle. This lack of data complicates the interpretation of our findings, and we hope that the extensive data obtained in this study will encourage researchers from various fields to explore this issue further. We also remind readers that the number of known spiral galaxies at high redshifts is still relatively small. For instance, \citet{Kuhn2023} reports 216 galaxies at $0.5 \leq z \leq 4$ classified as spirals, with unanimous classification for only 108 of them. Not all of these galaxies are likely suitable for decomposition, as a clearly defined spiral structure is required for decomposition to be feasible. Therefore, the 33 galaxies from the CEERS and JADES surveys represent a significant fraction of spiral galaxies at $z > 1$ for which decomposition is possible. One might even say that our study has approached the contemporary limit of what is currently achievable with existing observational data. Of course, one can expect that in the coming years JWST will provide us with even deeper and complete data. For example, recently~\citet{Wang2024} found an extremely large and evolved spiral galaxy at $z = 3.25$ in the JWST field.
	
\begin{figure}[!ht]
		\centering
		\includegraphics[width=0.99\linewidth]{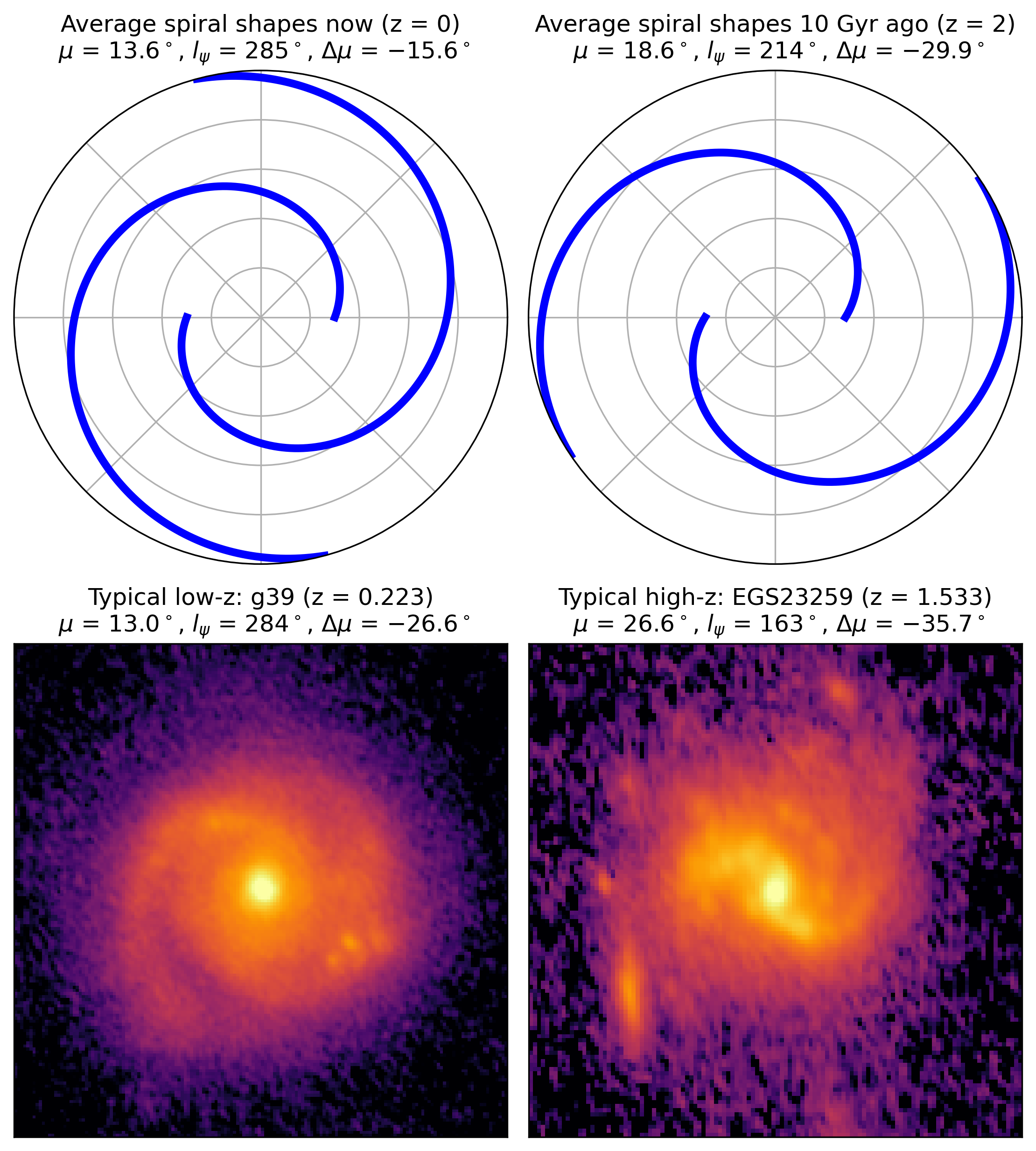}
		\caption{Top row: schematic representation of typical spiral arm shapes at the present time (at $z = 0$, left) and at a lookback time of 10 Gyr (corresponding to $z = 2$, right). For simplicity, we illustrate the case of a symmetric two-armed spiral structure. Bottom row: images of real galaxies, with g39 as an example of a low-$z$ galaxy (left) and EGS23259 in the F200W filter as an example of a high-$z$ galaxy (right).}
		\label{fig:typical_shapes}
	\end{figure}

	The observed variations in pitch angle $\mu$ and azimuthal length $l_\psi$ --- both parameters that define the general shape of a spiral arm --- with redshift $z$ or lookback time $t_L$ are among the most remarkable findings of our study (Sections~\ref{sec:pitch},~\ref{sec:length}). Having ruled out, to a certain degree, all observational effects (Sections~\ref{sec:band-shifting} and~\ref{sec:validation}), we conclude that the observed variations reflect the genuine evolution of these parameters on cosmological timescales (at least for bright galaxies from our sample). Specifically, spiral arms become more tightly wound, and their length increases with time, which, when combined, suggests a gradual winding up of spiral arms over time. We also measured another parameter that defines the shape of a spiral arm, $\Delta \mu$ (Section~\ref{sec:variations}). While the overall trend for $\Delta \mu$ is inconclusive, it nonetheless shows changes over time. To summarise our measurements and present them collectively, we have schematically and simplistically illustrated typical spiral arm shapes for the present (at $z \approx 0$) and 10 Gyr ago (at $z \approx 2$) according to our results, in Figure~\ref{fig:typical_shapes}. We also provide examples of galaxies with spiral arms that somewhat resemble these schematic shapes.
	
	What might be the possible reasons for such variations? Tidal and transient spiral patterns in individual galaxies are known to evolve over time, decreasing their pitch angle and increasing their azimuthal length (see, e.g., \citealp{Grand2013}). However, as \citet{Reshetnikov2023} argues, this is not an appropriate explanation for our observations, as these variations occur on shorter timescales and with larger amplitudes. The constant formation of tidal and transient arms is insufficient to account for the observed variations over cosmological timescales.
	
	\citet{Reshetnikov2023} also notes that galaxies become more bulge-dominated over cosmological timescales \citep{Bruce2014}, and that galaxies with more massive bulges and higher $B/T$ tend to have lower pitch angles (see, e.g., \citealp{Yu2019}). It is also observed that more massive and more luminous galaxies generally have lower pitch angles. While we do not intend to challenge these well-established observational facts, they do not align with our data. In our sample, distant galaxies are more luminous (see Figure~\ref{fig:basic_data}) and this also leads to higher $B/T$ for galaxies with higher $z$ (see Figure~\ref{fig:bulge_relations}) due to selection effects\footnote{Note that this does not contradict the fact that $B/T$ increases with time; see Section~\ref{sec:completeness}.}. Given these effects, one would expect that more distant galaxies in our sample would have lower pitch angles since they are more luminous and more bulge-dominated. However, we observe the opposite trend: pitch angles increase with $t_L$ in our data, despite the variations in $B/T$ and luminosity.
	
	Therefore, the observed trend is likely connected to other aspects of galaxy evolution and requires a different explanation. As the short-term evolution of pitch angles can depend on the disc rotational properties~\citep{Pringle2019}, and we are aware of the secular evolution of discs (see, e.g. the review of~\citealp{Sellwood2014}), this phenomenon can possibly be key to explaining the evolution of spiral structure. In this case, observed pitch angles should depend on how evolved the galaxy disc is rather than the age of the Universe, and as a possible way to examine this, one can study spiral arm properties in strongly evolved galaxies in the early Universe, such as the one discovered by~\citet{Wang2024} at $z = 3.25$.
	
	Regarding pitch angle variation along the spiral arm ($\Delta \mu$, Section~\ref{sec:variations}), we have already noted that the degree of variation changes over time. More importantly, we show that spiral arms tend to have a pitch angle that decreases from the beginning to the end, with an average overall decrease of $-17.3$ degrees. A similar observation was made for local galaxies by \citet{Savchenko2013}, who found that the majority exhibit decreasing pitch angles. Despite this, the logarithmic spiral with a constant pitch angle remains the most commonly used functional form to describe a spiral arm, mainly due to the simplicity of this function (see, for example,~\citealp{Lingard2020, Sonnenfeld2022}). Even though it is known that the pitch angle can vary along the length of the spiral arm, logarithmic spirals are often considered suitable to describe the ``average'' shape of spiral arms, particularly for simplification in theoretical works. However, this approach not only oversimplifies but also introduces a bias, as it fails to account for the typical decrease in pitch angle at the outer parts of a spiral arm (Chugunov \& Marchuk, in prep.).
	
	We applied the Pringle–Dobbs test \citep{Pringle2019}, see Section~\ref{sec:Pringle-Dobbs}, and found that the $\cot \mu$ distribution varies at different $t_L$, being most uniform at $t_L < 3$ Gyr and also showing a high degree of uniformity at $t_L \geq 8.5$ Gyr. Following \citet{Reshetnikov2023}, we interpret this as an indication of a change in the primary mechanism of spiral arm formation over time. Our proposed scenario is speculative and requires more evidence to draw a solid conclusion.
	
	Other measured parameters could potentially help differentiate between various spiral arm formation mechanisms. In particular, it would be informative to determine whether $\Delta \mu$ is connected with the mechanism of spiral arm formation. Hypothetically, if tidal arms were found to have a large negative $\Delta \mu$ on average, while density waves and transient spirals were closer to a logarithmic spiral shape, this would support our hypothesis, as $\Delta \mu$ deviates most significantly from zero at $t_L \geq 8.5$ Gyr.
	
	We emphasise that studying the evolution of spiral structures over cosmic time is complicated by band-shifting effects (Sections~\ref{sec:band-shifting},~\ref{sec:discerning}). Band-shifting effects are well known to be a significant factor in the study of distant galaxies, which is why the concept of K-corrections is widely used and continuously improved \citep{Hogg2002, Fielder2023}. However, the dependence of more specific structural parameters on wavelength, known as ``morphological K-correction'', is much less well established. Studies that use multiwavelength data to determine the basic relationships of structural parameters with wavelength are still in high demand; some recent examples include \citet{Buzzo2021}, \citet{Gong2023}, and \citet{Menendez-Delmestre2024}.
	
	Regarding spiral arms, there are few studies that examine how their properties depend on wavelength. The potential dependence of pitch angle on wavelength has garnered some attention because it serves as an observational test for density wave theory \citep{Yu2018b}, but other parameters remain largely unexplored. The rare examples of works where parameters of spiral structure are consistently measured across multiple bands are \citet{Kendall2011, Yu2018b, Savchenko2020} and our \citelinktext{Marchuk2024b}{Paper II}. 
	
	We found that band-shifting effects for spiral arms can be much stronger than evolutionary effects, as seen in the case of their width. To distinguish between these effects, a proper multiwavelength analysis is the most straightforward approach. However, we discovered that parameters do not change with $\lambda_\text{rf}$ beyond 1--1.5 $\mu$m (and at least up to $2.2 \mu$m), consistent with \citet{Ren2024}. Given this, single-filter data could potentially be used if the rest-frame wavelength of the filter falls within this range for the specified redshift range. Specifically, for the F444W NIRCam filter, with a pivot wavelength of approximately 4.4 $\mu$m, band-shifting effects will be minimal in the redshift range of $z\approx 1$ to $z\approx$ 2--3.
	
	Focusing on spiral structure asymmetry (Section~\ref{sec:asymmetry}), we have already noted that spiral structures become more asymmetric over time and less asymmetric at longer wavelengths. Notably, the average spiral arm asymmetry index is relatively high, at 0.44 for two-armed galaxies. This essentially means that 44\% of the luminosity of the spiral arms is not mirrored by the luminosity on the opposite side of the disc. This finding does not align with the expectation that spiral structures, particularly in grand-design galaxies (which are typically two-armed), are relatively symmetric; it is easy to see visually on images in Figure~\ref{fig:COSMOS_mosaic_1}.
	
	Furthermore, this presents a potential issue for any Fourier-based analysis, where the galaxy image is treated as a sum of symmetric modes. If 44\% of the total spiral arm flux is asymmetric, this asymmetry could lead to increased amplitudes of unexpected modes, complicating the interpretation of such analyses. This fact emphasises the importance of decomposition approach, which was developed in \citelinktext{Chugunov2024}{Paper I} \& \citelinktext{Marchuk2024b}{II}, and improved in this work.
	
	The overall shape of spiral arms, which we have discussed earlier, could also be a subject of such a study. To the best of our knowledge, most studies that have measured spiral arm parameters have focused on obtaining general characteristics rather than determining the exact functional form of their light distribution. We will pay attention to this question in our future work (Chugunov \& Marchuk, in prep.).
	
	\section{Conclusions}
	\label{sec:conclusions}
	
	Here, we summarise the main results of our study:
	
	\begin{enumerate}
		\item For the first time, we have performed photometric decomposition with spiral arms of 159 remote spiral galaxies at $0.1 \leq z \leq 3.3$, observed by HST and JWST. The examined sample is among the biggest compared to previous works on decomposition with spiral arms.
		\item We have confirmed our previous results concerning the change of ``classical'' components parameters after the addition of spiral arms to the model (see Figure~\ref{fig:classic_comps}). In particular, disc central surface brightness drops by 0.55 mag/arcsec$^{-2}$ on average, and bulge-to-total ratio increases by a factor of 76\%.
		\item We measured that pitch angle decreases over time with a rate of 0.5~deg/Gyr (see Figure~\ref{fig:mu-time}). This confirms the earlier results of~\citet{Reshetnikov2022, Reshetnikov2023} with a different method and also extends them, showing that this trend exists up to $t_L \approx 11$~Gyr ($z \approx 2.5$).
		\item We applied the Pringle--Dobbs test~\citep{Pringle2019} for different lookback time bins (see Figure~\ref{fig:Pringle-Dobbs}). The result is inconclusive; however, it suggests that the formation mechanism of spiral arms may change over time.
		\item We measured how pitch angles vary from the beginning to the end of a spiral arm. Similar to local spiral galaxies, the majority of objects in our sample show pitch angles that decrease toward the end of the spiral arm (see Figure~\ref{fig:dmu-time}). This decrease in pitch angle along the arm becomes much more pronounced for galaxies at $t_L > 8.5$ Gyr.
		\item We found that the azimuthal length of spiral arms increases over time at a rate of 7 deg/Gyr (see Figure~\ref{fig:length-time}), with 5 degrees per Gyr of this increase not being attributable to changes in resolution at higher $z$ (see Figure~\ref{fig:length-time-r25}). Combined with the decrease in pitch angles over time, this suggests that spiral arms are winding up as time progresses.
		\item We showed that band-shifting effects impact some spiral arm parameters mostly at wavelengths $\lambda_\text{rf}$ smaller than 1.0--1.5 $\mu$m, consistent with~\citet{Ren2024} --- see Figure~\ref{fig:params-lambda}.
		\item We observe that spiral-to-total luminosity ratio $S/T$ slightly increases with lookback time (0.08 per 10 Gyr). However, this appears to be merely a selection effect. This value also decreases towards longer wavelengths ($-0.03$ per 1~$\mu$m, see Figure~\ref{fig:ST-time-lambda}).
		\item We showed that the relative arm width $w$ normalised to the disc exponential scale length $h$ increases slightly with $t_L$ at a rate of $0.11 \times h$ per 10 Gyr but is strongly influenced by the rest-frame wavelength: $w$ increasing by $0.37 \times h$ per 1~$\mu$m (see Figure~\ref{fig:wh-time-lambda}).
		\item We found that the relative arm extent (the spiral end radius $r_\text{end}$ normalised to $h$) changes with time only weakly, if at all. The variation with rest-frame wavelength is moderate (see Figure~\ref{fig:rendh-time-lambda}).
		\item We measured the asymmetry index of spiral structure and found that it increases significantly with lookback time at a rate of 0.42 per 10 Gyr and decreases with wavelength at a rate of $-0.17$/$\mu$m. Overall, the spiral structure is fairly asymmetric, with the average asymmetry index for two-armed galaxies being 0.44 (see Figure~\ref{fig:asymm-time-lambda}).
		\item We performed a number of tests to ensure that our results concerning the change of parameters with $z$ reflect physical changes of spiral arms rather than show observational effects (see Section~\ref{sec:validation}). We found that, at least qualitatively, the results remain the same when observational effects are excluded.
	\end{enumerate}
	
	Overall, our study shows that spiral arms have become more tightly wound over the last 11~Gyr, and their spiral structures have become more symmetric (Figure~\ref{fig:typical_shapes}). Other parameters remain remarkably stable over time, although some are susceptible to apparent changes due to band-shifting effects. We expect that more spiral galaxies at high $z$ will be discovered in the near future, which will allow to confirm and strengthen the results obtained in this study. 
	
	\begin{acknowledgement}
		We thank the anonymous referee for their valuable comments and suggestions that helped us to improve the paper.
	\end{acknowledgement}
	
	\paragraph{Competing Interests}
	
	None.
	
	\paragraph{Data Availability Statement}
	
	Obtained decomposition results, including model files and obtained parameters, are available at \url{https://github.com/IVChugunov/Distant_spirals_decomposition}. Any other data underlying this paper will be provided on reasonable request to the corresponding author.
	
	\printendnotes
	
	\printbibliography
	
\end{document}